\documentclass{ar2e}
 
 \newcommand\ba{\begin{eqnarray}}
 \newcommand\ea{\end{eqnarray}}
 \newcommand\be{\begin{equation}}
 \newcommand\ee{\end{equation}}
 \newcommand\br{{\bf r}}

\def    \simlt  {\lower.5ex\hbox{$\; \buildrel < \over \sim \;$}}
\def    \simgt  {\lower.5ex\hbox{$\; \buildrel > \over \sim \;$}}
 
\begin{document}
 
 \jname{Annu. Rev. Astron. Astroph.}
 \jyear{2000}
 \jvol{1}
 \ARinfo{}
 
 \title{Chaos in the Solar System}
 
 \markboth{{Lecar,}{Franklin,}{Holman \&}{Murray}}{Chaos in the Solar System}
 
 \author{Myron Lecar, Fred A Franklin, Matthew J Holman
 \affiliation{Harvard-Smithsonian Center for Astrophysics, 60 Garden Street, Cambridge, Massachusetts 02138;
 e-mail: mlecar@cfa.harvard.edu}
 Norman W Murray
 \affiliation{Canadian Institute for Theoretical Astrophysics, McLennan
 Physical Labs, University of Toronto, 60 St. George Street, Toronto,
 Ontario M5S 1A7, Canada; e-mail: murray@cita.utoronto.ca }} 
 
 \begin{keywords}
 celestial mechanics, planetary dynamics, Kirkwood gaps
 \end{keywords}
 
 \begin{abstract}

 The physical basis of chaos in the solar system is now better
 understood: in all cases investigated so far,
 chaotic orbits result from overlapping resonances. 
 Perhaps the clearest examples are found in the asteroid belt.
 Overlapping resonances account for its Kirkwood gaps and were used to
 predict and find evidence for very narrow gaps 
 in the outer belt.  Further afield, about one new ``short-period''
 comet is discovered each year. They are believed to come from the
 ``Kuiper Belt'' (at 40~AU or more) via chaotic orbits produced by
 mean-motion and secular resonances with Neptune. Finally, the
 planetary system itself is not immune from chaos. In the inner solar
 system, overlapping secular resonances have been identified as the
 possible source of chaos.  For example,
 Mercury, in $10^{12}$~years, may suffer a close encounter
 with Venus or plunge into the Sun. In the outer solar system,
 three-body resonances have been identified as a source of chaos, but
 on an even longer time scale of $10^9$ times the age of the
 solar system.  On the human time scale, the planets do follow their
 orbits in a stately procession, and we can predict their trajectories
 for hundreds of thousands of years.  That is because the mavericks,
 with shorter instability times, have long since been ejected.  The
 solar system is not stable; it is just old!
 \end{abstract}
 
 \maketitle
 
 \section{INTRODUCTION}
 
 All known cases of chaos in the solar system are caused by overlapping
 resonances. Brian Marsden and I (ML), when we were students at Yale, 
 had frequent discussions with Dirk Brouwer about the significance of two 
 resonances that overlapped. The ``small divisors'' that occur in perturbation 
 theory at single resonances can be removed by a change of variables 
 introduced by Poincar\'e, which results in a resonance Hamiltonian similar to 
 that of a pendulum. No such device was ever discovered for two 
 overlapping resonances, and Brouwer sensed that there was something special 
 about that case. Because a single trajectory can be numerically integrated through 
 an overlapping resonance without apparent catasrophe, I argued forcefully
 (and incorrectly) that the difficulty with overlapping resonances might
 just be a defect in the perturbation theory. However, had we numerically
 integrated a clone, initially differing infinitesimally from the original 
 trajectory, the two would have separated exponentially.  In fact, that is the 
 definition of a chaotic orbit: exponential dependence on initial conditions. 
 
 In many cases we can estimate the Lyapunov Time (the e-folding time in the 
 above example) and even the Crossing Time (the time for a small body to 
 develop enough eccentricity to cross the orbit of the perturber). Both 
 times depend on the Stochasticity Parameter, which measures the extent of the 
 resonance overlap.
 
 There is evidence that, today, small bodies in the solar system (e.g.
 comets and asteroids) behave chaotically. Meteorites are thought to be
 fragments of asteroid collisions. The asteroid Vesta has a reflection 
 spectrum that resembles that of many meteorites. Every so often a meteorite 
 hits the Earth, so we have evidence, within the last hundred years,
  of chaotic behavior. 
 We have also found meteorites that originated on Mars. We believe they
 came from Mars because trapped bubbles of gas coincide with samples of the 
 Martian atmosphere. 
 
 Kirkwood (1867) noticed that the asteroid belt has ``gaps'' at resonances,
 i.e. at distances where the asteroidal periods are a rational fraction of
 Jupiter's period. There have been many attempts to explain these gaps on the 
 basis of a single resonance, but these attempts never produced gaps as 
 devoid of bodies as the observed ones. We now understand that bodies are 
 removed from regions where the overlap of two or more resonances induces chaos and 
 large excursions in the eccentricity. Wisdom (1982, 1983, 1985) first illustrated
 how chaos at the 3:1 resonance with Jupiter could result in sufficiently
 large eccentricity to allow an encounter with Mars or the Earth.
 Subsequent work showed that even collisions with Sun are a likely
 outcome (Ferraz-Mello \& Klafke 1991, Farinella et al 1994).  Chaotic
 dynamics in the asteroid belt will be reviewed in Section 3.  
 
 One to two ``short period'' comets (short means periods less than 200 years) 
 are discovered per year. Short period comets are confined to the ecliptic and 
 are believed to come from the Kuiper Belt, which is located about 40 AU from 
 the Sun, in neighborhood of Pluto. Their stability is also discussed
 in this review. 
 
 In contrast, long period comets are thought to come from the Oort Cloud
 at 20,000 AU. They are perturbed into the inner solar system by stellar 
 encounters or by the tidal field of the galaxy. These mechanisms differ 
 from those determining the dynamics of short period comets and are not
 reviewed here. 
 
 By now, the entire solar system exterior to Jupiter has been surveyed for 
 stability. Holman \& Wisdom (1993) found that all the unpopulated regions of 
 the solar system are unstable, on time scales that, in general, are much
 less than the age of the solar system. However, some comet orbits in the 
 Kuiper Belt are just now becoming perturbed into the inner solar
 system. The chaotic dynamics of comets in the outer solar system is
 reviewed in Section 4.  
 
 It is not too alarming that small bodies behave chaotically. Comets 
 and asteroids have individual masses less than 1/1000 that of the 
 Earth and, in total, make up much less than an Earth's mass. However, we 
 depend on the regularity of the planetary orbits. Could they be 
 chaotic? The answer for the known planets is yes---but on a long 
 time scale. 
 The planets have presumably followed their present orbits for much of
 the lifetime of the solar system.
 But for how 
 much longer? Are we in danger of losing a planet soon? Computers are just 
 now able to integrate the planets for the life time of the solar system so 
 we now have a preliminary exploration of this important problem. Here too
 chaos is induced at resonances, but evidence suggest that they are
 secular resonances that operate on long time scales. 
 
 Earlier, the stability of the solar system was studied by looking for
 terms in the semimajor axis that grew with the time, (secular terms), 
 or as the time multiplied by a periodic function of the time, (mixed
 secular terms). Now we know that instability comes from a chaotic
 growth of the eccentricity. The stability of the planets is reviewed 
 in Section 4. 
 
 For the reader who is intrigued, but new to this subject, we suggest a
 popular book called {\it Newton's Clock: Chaos in the Solar System} 
 by mathematician Ivars Peterson (1993). We refer all readers to the earlier 
 review by Duncan \& Quinn entitled ``The Long-Term Dynamical Evolution
 of the Solar System,'' which appeared in the 1993 edition of this Annual
 Review. In addition, the proceedings of the 1996 workshop on ``Chaos in
 Gravitational N-Body Systems'' have appeared as a book and in Volume 64 
 of {\it Celestial Mechanics and Dynamical Astronomy}. 

 H\'enon (1983) gives a lucid introduction to chaotic orbits and the
 ``surface of section'' technique.  Ott (1993) has written the
 standard text on chaos which covers a variety of physical problems.
 Sagdeev et al (1988) review other interesting applications of chaos
 including turbulence.
 
 \section{CHAOS AND CELESTIAL MECHANICS}
 
 The rigorous condition for a mechanical system to be stable for all 
 time is that there exists an ``integral'' (a conserved quantity) for each
 degree of freedom. The Sun and one planet is just such a system (e.g.
 the Kepler Problem). Fortunately, because the Sun is 1000 times
 more massive than Jupiter and the rest of the planets add up to less 
 than a Jupiter mass, treating the planets as independent two-body
 problems is an excellent starting approximation. Using their ``Keplerian''
 orbits, one can calculate a first approximation to their
 forces on each other. Successive iterations of that procedure, carried
 out with great sophistication, are the techniques called celestial 
 mechanics. The classic text on celestial mechanics was written by
 Brouwer \& Clemence (1961). See {\it Solar System Dynamics} by Murray
 \& Dermott (1999) for a recent treatment. 
 
 Much of the history of celestial mechanics has involved the search for
 integrals of motion. The search was doomed to fail; eventually Poincar\'e 
 proved that there was no analytic integral for the problem of the Sun 
 and two planets.  However, there do exist nonanalytic integrals. Their discovery
 culminated in the Kolmogorov-Arnold-Moser (KAM) theorem, a fundamental result in the mathematics
 of chaos. The theorem guarantees the existence of ``invariant curves''
 (i.e. other integrals) as long as the perturbations are not too large 
 and the coupling is not too near any resonance. Understanding the 
 exact meaning of the word {\it near} was crucial, because resonances (like 
 the rational numbers) are dense. This theorem is discussed in {\it The
 Transition to Chaos} by Reichl (1992) and in {\it Regular and Chaotic
 Dynamics} by Lichtenberg \& Lieberman (1992) (our recommended text). For the more 
 mathematically minded, there is also a set of lectures by Moser called {\it Stable and Random Motions in Dynamical Systems} (1973) and 
 {\it Mathematical Methods of Classical Mechanics} by Arnold (1978).

 Although the KAM theorem is of fundamental importance for the mathematical
 structure of chaos, the strict conditions of the theorem are
 satisfied in the solar system. In what follows, we will be concerned with
 orbits that are not covered by the KAM theorem.
 
 All of the analytic work described in this review relies on
 perturbation theory, exploiting the fact that the planetary masses are
 small compared with the mass of the Sun. A brief outline of the approach 
 follows.
 
The equation of motion of a planet orbiting a star accompanied by a
 second planet is
 \be \label{three_body} 
 {d^2\br_1\over dt^2}+{\cal G}(M_\odot+M_1){\br_1\over r_1^3}=
 {\bf\nabla}_1 R_{1,2},
 \ee 
 where
 \be 
 R_{1,2}={\cal G}M_2
 \left[{1\over r_{1,2}}-{\br_1\cdot\br_2\over r_2^3}\right]
 \ee 
 Here $r_1$ is the distance between the Sun and planet $1$, and
 $r_{1,2}\equiv\sqrt{(\br_1-\br_2)^2}$ is the distance between the
 planets. The quantity $R$, which is the negative of the planetary
 potential, is known as the disturbing function; it describes the
 disturbances of the planet's elliptical orbit produced by the other
 planet. The second term in the square brackets arises from the
 noninertial nature of the coordinate system employed and is known as
 the ``indirect'' term. It occurs because the traditional coordinate
 system takes the position of the sun as the origin. It is not generally
 relevant to the chaotic behavior of bodies in the solar system, so we
 ignore it henceforth.
 
 The next step is to expand the disturbing function using the
 expressions for $r_i$ and $\theta_i$ found by solving the Kepler
 problem. This rather daunting task has been performed by a number of
 authors (Peirce 1849, Le Verrier 1855, Murray \& Harper 1993). The
 general form is
 \ba \label{expansion} 
 R_{1,2}&=&{{\cal G}M_2\over a_2}\sum_{\bf j} \phi_{\bf
 j}(a_1,a_2)e_1^{|j_3|}e_2^{|j_4|}i_1^{|j_5|}i_2^{|j_6|} \nonumber\\
 &&\times\cos\left[j_1\lambda_1+j_2\lambda_2+j_3\varpi_1+j_4\varpi_2
 	+j_5\Omega_1+j_6\Omega_2\right].
 \ea 
 In this expression the $j_i$ are positive and negative integers. The angles in the argument of the cosine are the mean, 
 apsidal, and nodal longitudes and are measured from the x-axis. If we
 rotate the coordinate system, the disturbing 
 function, which is proportional to a physical quantity (the force), 
 cannot change. This implies that $\sum_ij_i=0$. We have kept only the
 lowest order terms in the sum; for a given $|j_i|$ terms proportional to
 $e^{|j_i|+1}$ or larger powers will also appear. Each cosine represents a
 resonance; the effects of these resonances constitute the subject of
 this review.
 
 We can use this expression to find the effect of one planet on
 another. Hamilton's formulation of mechanics offers the easiest way to
 proceed.  Using action-angle variables for the two body
 problem, the quantities $\lambda$, $-\varpi$, and $-\Omega$ are
 appropriate angles. The corresponding actions are simple functions of
 $a$, $e$, and $i$: 
 $L \equiv\sqrt{{\cal G}Ma}$, $G\equiv\sqrt{{\cal G}Ma}[1-\sqrt{(1-e^2)}]\approx (1/2)e^2L$, and 
 $H\equiv\sqrt{{\cal G}Ma(1-e^2)}(1-\cos i)\approx (1/2)i^2L$, where we
 have distinguished the gravitational constant, ${\cal G}$, from the
 momentum, $G$. In these variables the Kepler Hamiltonian is ${\cal H}=({\cal G}M)/2L^2$. Because
 we are using action-angle variables, none of the angles appear. This
 tells us immediately that the motion takes place on three-dimensional
 surfaces in phase space, defined by $L$, $G$ and $H$ held
 constant. Topologically, this is a three-torus. The surprising thing is
 that $G$ and $H$ do not appear in the Hamiltonian, so that $\varpi$
 and $\Omega$ are also constant. In a generic three degree of freedom
 Hamiltonian one would expect all three actions to appear explicitly,
 leading to three non-constant angles. The Kepler problem is
 degenerate, 
 as is reflected in the fact that $H$ and $G$ do not appear in 
 ${\cal H}$. In this case the motion takes place on a one-torus, or 
 circle, in phase space.
 
 To find the variation in $a$, for example, note that
 \be 
 {dL_1\over dt}={1\over2}{L_1\over a_1}{da_1\over dt},
 \ee 
 so
 \be \label{lagrange} 
 {da_1\over dt}=2{a_1\over L_1}{\partial R\over\partial\lambda_1}.
 \ee 
 The presence of planet 2 forces periodic variations in the semimajor
 axis of planet 1. For example, suppose we pick a term of the form
 \be 
 \phi_{2,-5,3,0,0,0}(a_1,a_2)e_1^3\cos[2\lambda_1-5\lambda_2+3\varpi_1].
 \ee 
 The equation for $a_1$ becomes
 \be \label{mean_motion} 
 {da_1\over dt}=-4(a_1n_1){a_1\over a_2}{M_2\over M}\phi_{2,-5,3,0,0,0}(a_1,a_2)e_1^3
 \sin[2\lambda_1-5\lambda_2+3\varpi_1]+O(e)^4.
 \ee 
 Similar expressions can be derived for all the other orbital
 elements. 
 
 With these expressions, or similar ones provided by Lagrange, it
 appeared to be a simple matter to integrate the equations of
 motion. However, early efforts to do so revealed difficulties.
 The problem can be seen by integrating equation (\ref{mean_motion}).
 To first order in $M_2$ we find
 \be \label{a_52} 
 a_1(t)-a_1(0)=a_1{a_1\over a_2}{M_2\over M}{n_1\over 
 2n_1-5n_2+3\dot\varpi_1}\phi_{2,-5,3,0,0,0}(a_1,a_2)e_1^3
 \cos[2\lambda_1-5\lambda_2+3\varpi_1]+O(e)^4.
 \ee 
 The denominator $2n_1-5n_2+3\dot\varpi_1$, and similar denominators
 that arise when all the terms in the disturbing function are
 considered, is the source of the difficulty. Poincar\'e (1993) pointed out that
 for integers $j_1$ and $j_2$ (2 and 5 in our example) the
 denominator becomes arbitrarily small. He went on to show that, in
 spite of the fact that terms with large $j_1$ and $j_2$ tend to carry
 large powers of the eccentricities, the sums used to define the
 hoped-for analytic solutions diverged. 
 
 The locations in phase space (essentially along the $a$ axis) where
 the denominator vanishes are known as resonances. Poincar\'e noted that
 in the immediate vicinity of a resonance the motion was very
 complicated. Later work, particularly that of Chirikov (1979), showed that
 the complicated motion, dubbed chaos, occupied a large fraction of the
 phase space near the resonance when two neighboring resonances
 overlapped. This result is essentially an anti-KAM theorem, in that it
 specifies where KAM tori are absent. It provides the underpinings of
 much of the work reviewed in this article.
 
 \subsection{The Origin of Chaos: Overlapping ``Mean-Motion'' Resonances}
 
 The chaos generated at overlapping resonances was first studied, in the
 astronomical context, by Wisdom (1980), who calculated the width and
 extent of overlap of adjacent first-order resonances in the circular restricted 
 three-body problem and found that they overlap to a distance given
 by $\delta a/a \sim 1.3 \mu^{2/7}$, where $\mu$ is the mass ratio of 
 the planet to the central star. The resonance overlap criterion for
 chaos was developed by Chirikov (1979). The width of the first-order
 resonances, for zero eccentricity, in the restricted problem was also 
 derived by Franklin et al (1984), from which we find
 that the 2:1 mean-motion resonance extends, in semimajor axis (with
 $a_{Jupiter} = 1$) from 0.621 to 0.639, the 3:2 from 0.749 to 0.778, the 4:3 
 from 0.808 to 0.843, the 5:4 from 0.841 to 0.883, and the 6:5 from 0.863 
 to 0.908. The 4:3 overlaps with the 5:4 and all adjacent resonances closer
 to Jupiter overlap. The 4:3 resonance is at 0.825; the Wisdom formula
 predicts onset of chaos for $a > 0.82$. This region where mean-motion resonances overlap is where
 the relation between the Lyapunov time, $T_c$, and the time for a close
 encounter with the perturber, $T_c$, applies. This relation, found
 empirically by 
 Lecar et al (1992), predicts that $T_c$, is proportional 
 to $T_l^{1.75}$. 
 
 Murray \& Holman (1997) have explained this relation in terms of the
 Stochasity Parameter, $K = (\pi \Delta \Lambda/\delta\Lambda)^2$, where 
 $\Delta\Lambda$ is the width of the resonance and $\delta\Lambda$ is the 
 separation between the resonances. If $\Delta K = K - K_c$, where $K_c$
 is the critical value ($\sim 1$, corresponding to the start of
 overlap), then they showed that $T_l \sim \Delta K^{-1.65}$ and $T_c
 \sim \Delta K^{-2.65}$, so $T_c \sim {T_l}^{1.6}$.  This holds in the
 region of overlapping first-order resonances (the $\mu^{2/7}$
 region). Orbits in this region show a range of more than three orders
 of magnitude in $T_c$.  
 However, outside that region, they showed that the relation was in error
 by a factor of 10 at the 5:3 resonance (a second-order resonance), and by
 a factor of 100 at the 7:4 resonance (a third-order resonance). They also
 integrated 10 ``clones'' of Helga, an asteroid at the 12:7 resonance (a
 fifth-order resonance). Five of the clones had Lyapunov times ranging from
 6,000 to 13,000 years. They encountered Jupiter in 1--4~Gyr (Gyr = $10^9$ years).
 In this case the relation predicted $T_c\sim 6$~Myr (Myr = $10^6$ years)---too low
 by a factor of 1,000. Murray \& Holman (1997) showed that the mechanism for chaos in these
 higher order resonances was an overlap of the ``subresonances,'' and
 that diffusion between overlapping subresonances in the same
 mean-motion resonance is slower than diffusion between overlapping
 mean-motion resonances.
 
 It is worth noting that higher order resonances become very narrow
 for zero eccentricity.  For example, from Franklin et al (1984), 
 the width of first-order resonances is
 proportional to $\mu^{2/3}$ for zero eccentricity.  The corresponding
 widths of second-order resonances is proportional to $\mu$ and the
 width of third-order resonances is proportional to $\mu^2$. Thus,
 second and higher order resonances are too narrow to overlap each
 other at low eccentricity.  However, higher order resonances can
 occur in the wings of first-order resonances. 

 \section{DYNAMICS IN THE ASTEROID BELT}
 
 Understanding the distribution of the asteroids as a function of
 their semimajor axes, and particularly where their mean-motions are
 commensurate with Jupiter's, provided dynamicists an intriguing
 puzzle for over 130 years---all the more so because these (Kirkwood) gaps
 occur at most mean-motion ratios, $n_A/n_{Jup} = p/q$, but a
 concentration of bodies at two others. Progress on this classic problem
 has been striking over the past 20 years. In a broad sense, it has been
 solved: We can identify the sources of orbital instability (or their
 absence) and the nature of their consequences and also have a good idea of 
 some of the time scales involved. Numerical and analytic studies both 
 have contributed extensively. Although several related dynamical
 processes have been---and still are---working to produce gaps in the 
 asteroid distribution, the most significant ones can all be linked to 
 the solar system's present environment. Carving gaps may in 
 some cases require upwards of a billion years, but it can probably be 
 done without requiring cosmogonic explanations; i.e. calling on
 processes that occured in the primordial or developing solar system.
 
 The paper that ignited the modern era of work on the Kirkwood
 problem was Jack Wisdom's (1982) first contribution to the study
 of the 3:1 mean-motion resonance at $a$ = 2.50~AU. 
 His startling results 
 showed that an orbit at this resonance could remain quiescent, with a 
 low eccentricity, $e < 0.1$, for more than 100,000 years but also show 
 occasional surges lasting about 10,000 years that would lift e to a
 maximum value of about 0.35. Such a value is just sufficient to allow a
 crossing of Mars' orbit, resulting in an eventual collision or a close
 encounter. Orbital computations as long as a million years were a
 rarity 20 years ago, and Wisdom's novel approach was to develop a mapping 
 of the planar elliptic three-body problem that relied on two efficient
 techniques. The first approximated the short-period terms (i.e. ones 
 characteristically arising during an orbital period) by a series of
 delta functions. The second averaged the Hamiltonian, expanded to second
 order in eccentricity, over the longer but still relatively short-term
 angular variable that librates at the 3:1 resonance.
 
 In two following papers, Wisdom (1983, 1985) first used direct
 numerical
 integrations to verify the presence of the $e_{max}$ peaks of 0.35 and, by
 including a third dimension, then showed that $e_{max}$ could rise to 
 e = 0.6, a value that also included Earth-crossing trajectories. At the
 same time he calculated the extent of the chaotic zone, showing that it 
 closely matched the observed 3:1 gap width. The excitement generated by 
 these results echoed widely: A straightforward dynamical process that 
 ({\it a}) could open a gap at one resonance, ({\it b}) in principle might be generalized 
 to account for other Kirkwood gaps, and as an added bonus,  ({\it c}) could
 deliver asteroidal fragments into the inner solar system as meteorites,
 had at long last been identified. Later work by Ferraz-Mello \& Klafke
  (1991) and Farinella et al (1994) showed that the chaotic zone at low e is linked, even in the 
 planar elliptic three-body problem, to one with $e > 0.6$ so that 
 $e \rightarrow 1$ can occur.

 The panels of Figure~\ref{FigA} provide a quick insight into
 the chaotic behavior that Wisdom discovered at 3:1. Figure~\ref{FigA}{\it a}
 plots the motion of both Jupiter's apsidal line, $\varpi_J$, and that of
 a low eccentricity body [$e_o = 0.05$] in the 3:1 resonance, and
 Figure~\ref{FigA}{\it b} shows that the eccentricity surges occur when the two
 $\varpi$'s are approximately equal---in fact, $\varpi_A > \varpi_J$ 
 corresponds to the rise in e. The equality of two apsidal or nodal rates
 is referred to in solar system studies as an example of secular 
 resonance. Their importance at certain locations in the asteroid belt
 has long been recognized (cf Brouwer \& Clemence 1961), but only in the
 past decade has their role within mean-motion resonances been
 appreciated. Perturbations arising at mean-motion resonances will
 markedly effect the elements of bodies lying in them---yielding
 a very broad range of apsidal and nodal rates that are functions of e
 and i. The apsidal and nodal motion of the Jupiter-Saturn system is
 defined to high precision by two apsidal and one nodal terms that have been 
 labeled $\nu_5$, $\nu_6$, and $\nu_{16}$ by Williams (1969).
 Figure~\ref{FigA}$a$ shows that the apsidal motion of a body with 
 $a_0 =0.481$ and $e_0 = 0.05$, which librates in the 3:1 resonance,
 will intermittently also resonate with the frequency of $\nu_5$ and
 possibly $\nu_6$ as well.
 
 The challenge to map the locations and limits of the secular
 resonances that lie within the confines of many significant mean-motion
 resonances has been met on theoretical grounds in papers by Moons \& 
 Morbidelli (1995 and its references).  Although
 this work concentrates on the planar case in which only $\nu_5$ and 
 $\nu_6$ are present, it contains the important result that, in the higher 
 order resonances, 3:1, 5:2, and 7:3, the $\nu_5$ and $\nu_6$ secular
 resonances exist over a very wide and overlapping range of a and e---to
 such an extent that a condition of widespread chaos is present inside 
 these three mean-motion resonances. Figure~\ref{FigB} is a sample of
 their work for the 5:2 resonance.
 
 The picture developed by Moons \& Morbidelli (1995) provided one reason 
 that led Gladman et al (1997) to study the fates of a large number of
 bodies placed in various mean-motion resonances. In a real sense their 
 paper represents an elaboration and even a culmination of Wisdom's
 original suggestion that marked eccentricity increases are responsible
 for the Kirkwood gaps. Some eccentricity increases beyond the value of
 0.35 found by Wisdom had already been noted, but Gladman et al provided
 accurate statistics by integrating more than 1000 bodies sprinkled
 throughout 3:1, about 450 in 5:2, and 150 each in 7:3, 8:3, and 9:4.
 Their study quantitatively describes the dynamical transfer process
 noted earlier by Wisdom (1983)---namely that gravitational encounters
 even with the terrestrial planets can provide sufficient
 energy changes to move bodies from regular to chaotic zones and even
 from one resonance to another. We can now legitimately claim that the 
 development of a gap [cf Figure~\ref{FigC}] at 3:1 is inevitable, though
 some details are complex and different time scales are followed. The
 next two paragraphs provide an outline of the process. 
 
 First, for bodies once in 3:1 with $e < 0.25$, the effect of imbedded 
 secular resonances, principally $\nu_6$, will drive e's of any and all
 bodies toward unity, leading most likely to solar impacts in times of a few million 
 years. In the survey, this was the fate of about 70\% of the initial
 population. A quarter to a third of them directly impacted the Sun, whereas 
 the majority were gravitationally scattered by the Earth or Venus before 
 doing so. Most of the remaining 30\% moved in unstable orbits exterior to 
 Saturn. 
 
 Second, the same fate awaits bodies of any eccentricity whose
 critical arguments are either circulating or that have librations
 greater than about $50^\circ$. However, bodies with $e > 0.3$ and that show
 small librations have different outcomes. In preparing for this
 review, we identified orbits of bodies with $0.3 < e < 0.6$, whose moderate 
 librations, all $< 40^\circ$ qualified them as stable librating members
 of the 3:1 resonance. These orbits are regular; i.e. there is no sign of 
 any exponential growth in their angular orbital elements during integrations 
 lasting as long as $10^7$ yrs. Figure~\ref{FigD}$a-c$ is an example of
 one of them. Three (of three) remained in regular orbits until the
 integrations were terminated after 2 billion years. However, the 
 eccentricities of all three of these ``stable'' orbits regularly climbed at 
 least to 0.5, hence risking the gravitational encounters (or, less probably, 
 actual collisions) with Mars mentioned above. (Mars itself was not 
 included in these integrations.) Gladman et al found that 5\% of all
 bodies initially in 3:1 were ``extracted by Mars,'' meaning that their orbits
 were first perturbed by Mars by a sufficient amount that the final result 
 after subsequent encounters was most likely a solar impact or an orbit 
 beyond Saturn. This seems the certain fate of the otherwise 
 stable high eccentricity bodies---``otherwise'' meaning the case with only 
 Jupiter and Saturn present. Their result argues that this phase of
 the depopulation of 3:1 will require longer times, 10--100~Myr (with
 a tiny handful remaining after 100~Myr, but the eventual
 outcome is the same as that of their lower eccentricity neighbors. We
 can conclude that 3:1 is a resonance that has been emptied of any
 asteroids initially present by a natural, multi-stage dynamical process
 in which all planets, Venus through Saturn, have contributed.
 
 The 5:2 resonance, although one order weaker than 3:1, behaves
 similarly but with some interesting differences. As was the case at
 3:1, values of the eccentricity set three regimes: ({\it a}) Orbits of low e
 are severely chaotic owing to the influence of $\nu_6$. This remark 
 applies to all bodies with $a \simeq a_{5:2} = 2.78$~AU and all $e < 0.2$. 
 Figure~\ref{FigE}{\it a-d} presents an example. ({\it b}) for $0.2 < e <
 0.40$, orbits are regular, provided that their libration amplitudes are 
 less than $11^\circ$ and ({\it c}) for all larger e's, orbits at least
 occasionally lie in secular resonance and are very chaotic with e's 
 reaching values $>0.7$. These remarks emphasize the resemblence to
 3:1, but there are two novel features for bodies in the second category. 
 
 First, their $e$ ranges generally do not include a close approach to Mars
 and second, as figures~\ref{FigE}{\it c,d} show, in just one of many cases, the
 presence of secular resonance does not always correspond to extreme
 chaos. Taken together, figures~\ref{FigE}{\it a-d} are examples of the 
 differing response of two bodies at the 5:2 mean-motion resonance to
 secular resonances. When the amplitude of $\varpi_A - \varpi_J$ is large
 or when there is coincidence of the periods of the $\nu_6$ term with
 the $\varpi_A$ oscillations, then chaos is severe and is measured by
 Lyapunov times of $\sim1000$~Jovian periods. By contrast, when the 
 oscillations of $\varpi_A$ no longer match the frequency of $\nu_6$ 
 and/or their amplitude is small the orbits remain regular. A tentative 
 conclusion is that $\nu_5$ is far weaker than $\nu_6$. Thanks to Moons 
 \& Morbidelli (1995) we have an accurate knowledge of the locations of 
 the secular resonances as well as the limits on the orbital elements over
 which they operate. What we lack is an evaluation of their strength
 ---say, a measure of the chaos they can produce within a mean-motion
 resonance. Murray \& Holman (1997) have developed the means to
 calculate Lyapunov times when mean-motion resonances overlap. A related
 formalism applying to the case when secular and mean-motion resonances
 overlap would be very valuable. 
 
 In their study Gladman et al found that the loss of bodies from
 5:2 closely resembled the depletion at 3:1; i.e. all objects were removed 
 within 100~Myr. Our suggestion is otherwise: We expect a
 number of librating bodies with $ e < 0.40$ to remain, unless a slow
 diffusion into the chaotic region has removed them all. However, there may
 not be a real conflict. Gladman et al were especially interested in the
 role of the 5:2 in delivering members of three nearby asteroid families
 into the inner solar system and hence chose initial velocity 
 distributions accordingly. Inclusion of librators in such a
 distribution is very unlikely. By contrast, their survey at 3:1,
 though it also followed the evolution of members of three families,
 introduced 1000 randomly selected orbits as well. The files of the
 Minor Center list some 30 possible candidates in the above eccentricity
 range at 5:2. Whether they do librate with small amplitudes and avoid
 secular effects is at present unknown. 
 
 Moons \& Morbidelli (1995) have also mapped the one (weaker)
 secular resonance, $\nu_5$, that lies within the 4:1 mean-motion
 resonance, but 4:1 itself is unimportant because its semimajor happens
 nearly to coincide with the locations of the far stronger $\nu_6$ and 
 $\nu_{16}$ secular resonances that, independently of any mean-motion
 resonance, are now known to be the agents defining the inner edge of the 
 asteroid belt at 2.1~AU.
 
 Both Moons \& Morbidelli (1995) and Gladman et al (1997) also examined
 the 7:3 mean-motion resonance. The former paper again shows that both 
 $\nu_5$ and $\nu_6$ are centrally situated within 7:3 and cover large
 portions of it, both as $e \rightarrow 0$ and at e as high as 0.65 as
 well. In preparing for this review, we randomly introduced 120 bodies into the resonance and found
 severe chaos ($T_l < 2000 P_J$) everywhere. Gladman et al (1997) found that
 depletion at 7:3 proceeds at a slower rate than at 5:2, with nearly
 one half 
 of its initial bodies surviving to their integration limit of 40~Myr. Our 
 results are roughly compatible with theirs but suggest a depletion rate 
 that is faster by a factor of at least two. In any event, as
 Figure~\ref{FigC} attests and both results predict, 7:3 should be a genuine,
 fully developed gap, not just a region of reduced population. 
 
 The fifth-order resonances, 8:3 (a = 2.71~AU) and 9:4 (a = 3.03~AU)
 were not examined in the otherwise comprehensive studies of Moons \& 
 Morbidelli so we have introduced about 50 bodies in each to have some
 idea where chaos is most important. The results contain few surprises:
 at 8:3 orbits with $e_p < 0.19$ are severely chaotic, with 
 $T_l$'s $< 2000 P_J$. In the range $0.19 < e_p < 0.35 $ they are quite
 regular, having $log T_l > 4.5 $ (Jovian periods). They become
 increasingly chaotic at higher $e$'s, a fact that is of little 
 importance because $e = 0.4$ leads to a Mars encounter. At 9:4 orbits
 are again very chaotic for $e_p < 0.2$ and much less so for 
 $0.2 < e_p < 0.32 $. Still, their average $T_l$'s are shorter by a
 factor of about three than those at 8:3. Eccentricities $> 0.32$ are again
 very chaotic, and those with $e > 0.45$ will be Mars crossers. Gladman et
 al (1997) found incomplete depletion at both 8:3 and 9:4, with about one half of the
 original population surviving after 40~Myr in the former and three quarters 
 after 120~Myr at the latter. All of these estimates are qualitatively in
 accord with the distribution shown in Figure~\ref{FigC}. [A likely
 interaction between 9:4 and 11:5 (3.08~AU) probably accounts for the
 extra width.] Longer surveys, with an eye to further quantifying
 diffusion rates into chaotic areas, are an important future project.
 
 \subsection{Chaos in the Outer Belt}  

 The 2:1 resonance at 3.28~AU divides the populous inner belt from the
 much less dense outer portion [cf Figure~\ref{FigC}]. Can it be that
 the entire outer belt is systematically more chaotic than the inner
 belt? The answer is certainly yes, and it
 has recently become possible to evaluate both its extent and severity 
 throughout all of the outer belt where Holman \& Murray (1996) have,
 among other results, shown that 22 of 25 outer belt minor planets
 have Lyapunov times shorter than $6000 P_J$. However, it is also important to
 stress that there is more chaos in the inner belt than what we have thus far mentioned. Independent studies by Murray et al
 (1998) and Nesvorn\'y \& Morbidelli (1998) point out that many three-body
 mean-motion resonances (cf Aksnes 1988), involving the longitudes of Saturn as 
 well as Jupiter and the asteroid, are a major source of
 chaos. Nesvorn\'y and Morbidelli (1998) trace the chaos in the orbits of about 250 of just the
 numbered minor planets to three-body resonances, finding Lyapunov times 
 that lie in the range of 1,000 to 10,000 $P_J$. As the authors demonstrate, these 
 resonances are vastly more dense in the outer belt, where their effect can 
 only be more destabilizing.

 Holman \& Murray (1996) and Murray \& Holman (1997) have studied 
 chaos in the outer belt by examining orbital behavior at a number 
 of high-order resonances (e.g. 12:7) where they have calculated both
 Lyapunov and diffusion time scales, analytically as well as numerically, 
 and some escape times for comparison. They associate chaos with the
 overlap between members of individual mean-motion resonances. Consider the 
 example of 12:7. In the planar case it is composed of six (sub)resonances, with various 
 multiples of the two apsidal longitudes, e.g.
 \be
 \sigma_k = 12 \lambda_J - 7 \lambda_A - k \varpi_A + (k-5)
 \varpi_J, \label{sigma}
 \ee 
 for $ k =
 0,1,...5 $ and where the $\lambda$'s and $\varpi$'s refer to mean and
 apsidal longitudes. Two components, defined e.g. by k and k+1 can
 overlap and Holman \& Murray (1996), having obtained their locations
 and widths, showed that their overlap will generate chaos. This approach
 to analyzing chaos within a given mean-motion resonance is
 the mathematical equivalent of the one used by Moons \& Morbidelli (1995),
 who consider the overlap between a resonance defined by $\sigma(k) $
 in Equation~\ref{sigma} and the secular one given by \be \sigma_{secular} \equiv
 \varpi_A - \varpi_J = [\sigma(k+1) - \sigma(k)]. \ee The latter is
 the approach taken by Wisdom in his analysis of the 3:1 resonance
 (Wisdom 1985). In this approach it is sometimes assumed that there
 is an adequate separation between the time scales associated with
 $\sigma(k)$ and $\sigma(secular)$ so that the action associated with
 $\sigma(k)$ is adiabatically preserved, except when the orbit is
 near the separatrix. However, these time scales are not
 adequately separated for most resonances in the asteroid belt.
 
 Figure~\ref{FigF} plots some of Holman \& Murray's results, indicating that
 ({\it a}) asteroids in the outer belt have already been ejected from
 resonances of order less than 4 (with the exception of 11:7), thus
 demonstrating that less well-defined Kirkwood gaps will also exist there; 
 ({\it b}) escape times from fifth-order resonances correspond roughly to the solar 
 system's age' and ({\it c}) objects in sixth-order resonances are ejected in times as long
 as $10^{11-12}$~years. We encounter again the problem discussed
 elsewhere in this review: Lyapunov times are measured in a few thousand 
 $P_J$, but at least some of the carefully evaluated escape times exceed
 the age of the solar system. In view of these long times for ejection,
 it seems likely that other processes have contributed to the removal of 
 some resonant and many nonresonant bodies from the extended outer belt.
 Holman \& Murray (1996), Liou \& Malhotra (1997), and Nesvorn\'y \&
 Ferraz-Mello (1997) have directed attention to planetary migration as a
 mechanism that will move various resonances into and through the outer
 belt, a process that can have further dynamical consequences, as the
 latter paper stresses, if there is also a change in the near 5:2
 commensurability in the mean motions of Jupiter and Saturn. Lecar \&
 Franklin (1997) and Franklin \& Lecar (2000) have quantitatively
 studied the sweeping of secular resonances associated with the decay 
 of the solar nebula through the asteroid belt. These papers show that
 this one mechanism can accomplish three desired ends: ({\it a}) remove an overwhelming 
 fraction of an initial population from the outer belt, ({\it b}) deplete the
 inner belt from a likely early value by a factor of about 1000 so as to 
 match present observations, and ({\it c}) generate a range of eccentricities that 
 are characteristic of the known minor planets. Their study also included
 gas drag on the asteroids. 
 
 \subsection{Behavior at First-Order Mean-Motion Resonances}        

 We turn at last to the behavior at two first-order resonances that 
 Morbidelli \& Moons once called the ``most mysterious ones,'' beginning with 
 a discussion of the reason(s) for the pronounced gap at the 2:1 resonance,
 $a_0 = 0.630$ and the concentration of minor planets at the
 3:2 resonance, $a_0 = 0.763$. Part but not all of this mystery has been
 dispelled as models have become more realistic. To be more precise, unless
 models include Jupiter and Saturn in eccentric precessing ellipses,
 they can not capture enough physics to account for the behavior at the
 first-order mean-motion resonances. The importance of secular resonance 
 in developing chaos within such higher order mean-motion resonances as 3:1 
 and 5:2 might lead to the expectation that they should also be of prime 
 importance here as well. However, a detailed mapping (Morbidelli \& Moons
  1993) shows that the role of $\nu_5$ and $\nu_6$ for generating chaos 
 in the heart of both the 2:1 and 3:2 resonances applies
 only for orbits having eccentricities $e > 0.45$ and $0.25$, respectively. 
 The reason is clear enough: At low e's, first order resonances drive
 pericentric longitudes, $\varpi_A$, much more rapidly than the motion
 of even the faster of the two principal terms (i.e. $\nu_6$) that measure 
 $\varpi_J$---by a factor of about 30 for orbits with $e$'s averaging about
 $0.15$. We must look elsewhere for other resonances whose overlap at 
 orbits of low e with the 2:1 resonance itself is the source of the chaos 
 shown in figure~\ref{FigG}{\it a}.
 
 The ones we are looking for involve commensurabilities between 
 multiples of the (resonant) libration and apsidal frequencies. The former
 are defined by the regular oscillations of the appropriate critical angle
 $\sigma$ similar to Equation~\ref{sigma}. Strictly periodic solutions have
 $\sigma = 0$; real bodies at mean-motion resonances show oscillations in
 $\sigma$ with well-defined periods and amplitudes that also correspond to
 regular variations in their orbital elements, especially semimajor axis
 and eccentricity. Periods of $\sigma$ increase with $e$, and at 2:1 and
 3:2, typically lie in the range of 15 to 75 $P_J$, where $P_J$, the orbital
 period of Jupiter, is 11.86~years. Apsidal periods also depend on e, from close 
 to $15 P_J$ for very small values, then increasing rapidly. Therefore, a 
 series of commensurabilities, starting near unity, which we label 
 $P_{apse}/P_{lib}$ in Figures~\ref{FigG} and \ref{FigH} must occur.
 
 Giffen (1973) first called attention to these commensurabilities,
 though their link to
 chaos was not stressed. Henrard and colleagues (cf Henrard and Lemaitre 1986, Lemaitre
 \& Henrard 1990 first recognized their importance and began to
 investigate them analytically, and Franklin (1994, 1996), numerically.
 Following Henrard we shall refer to the small integer commensurabilities of 
 $P_{apse}/P_{lib}$ as secondary resonances. We can explore their
 behavior and importance by considering the planar example shown in 
 Figure~\ref{FigG}{\it a--c}. The figure shows that they are indeed the source of 
 chaos within the 2:1 resonance by presenting three cases in which the two perturbers, 
 Jupiter and Saturn, move in planar precessing orbits with (case a) their present
 eccentricies ($e_J = 0.044 \pm 0.016$; $e_S = 0.047 \pm 0.035$), then with
 one half (case b) and finally one tenth (case c) of these values. This sequence
 helps verify the claim that secondary resonances are responsible for 
 chaos among (hypothetical) resonant asteroidal orbits of low
 eccentricity. [A slash, e.g. 5/3, denotes a secondary resonance,
 while a colon, e.g. 3:2, denotes a mean-motion resonance.]
 Figure~\ref{FigG}{\it c} shows that only the strongest secondary resonances are 
 present and their widths are very narrow, but in figure~\ref{FigG}{\it a}
 when e(J) and e(S) have normal values, higher order secondary resonances 
 [e.g. 5/3] have also been excited and/or first-order ones have broadened
 so that extensive overlap occurs.
 
 A quick interpretation of figure~\ref{FigG} might suggest 
 that secondary resonances are the source of chaos in the entire range, 
 $0 < e < 0.28$. If this is truly the case, then ones as high as 15/1 must
 still be contributing. Despite our present ignorance of their strengths,
 this in itself seems curious, though since $P_{apse}$ is a rapidly rising
 function of e(asteroid), whereas $P_{lib}$ is almost constant, it is also true
 that higher order resonances crowd closer and closer together. The fact
 that exactly at some weaker secondary resonances very regular orbits exist 
 is also curious, though helpful, as it speeds the process of locating them. 
 A supplementary explanation for chaos near $e = 0.2$ has been mentioned by 
 Morbidelli \& Moons (1993), who found that the $\nu_{16}$ secular resonance is 
 present in the limit as $i \rightarrow 0$ and should affect orbits with 
 $e = 0.21 +0.04/-0.02$. However, the calculations leading to 
 figure~\ref{FigG} are strictly limited to the planar approximation so that 
 the origin of chaos for $e_0 < 0.15$ is clearly the province of secondary 
 resonances, but to what e they extend is less clear. 
 
 Recently, Moons et al (1998) broadened this topic by mapping the
 locations of chaotic zones that arise from all secular and families of 
 secondary resonances at inclinations of 0, 10, 20, and $30^\circ$. [As
 Henrard (1990) noted, secondary resonances are not confined to
 the case defined by $\dot\sigma = \dot\varpi_A$, but can include linear 
 combinations of $\dot\varpi_A$ with the frequencies of planetary apsidal
 and nodal motion.] What is striking is the connected nature of the chaotic
 zones, extending through much of the area in the $a,e$ plane for all four
 $i$'s. Depending on the inclination, one to three islands are present that
 contain quite regular orbits, having $0.2 < e < 0.45$, many of which are not 
 at risk because their perihelia avoid crossing Mars' orbit. Moons et al 
 note that one of them is populated by five recently discovered minor planets, 
 but the others are seemingly empty.
 
 With the source of chaos identified, the question now turns to how the
 depletion of a hypothetical early population of bodies at 2:1 might have 
 proceeded. This topic rests on less firm ground. Two independent sets of 
 long-term integrations argue that objects with Lyapunov times as short
 as several hundred Jovian periods (a few thousand years) cannot 
 permanently exist at 2:1. Franklin (1996) found examples of escape for 
 single representative bodies placed in the 2/1, 3/1 and 5/1 secondary 
 resonances, all after times close to 800~Myr. Objects in much
 higher secondary resonances, consequently with Lyapunov times three orders
 longer, remained after 4~Gyr. These integrations included only 
 Jupiter and Saturn, which moved in planar precessing orbits. In five cases of escape 
 orbital eccentricities rose to values that guaranteed a crossing of Mars'
 orbit only a few million years prior to escape, leading to the conclusion
 that drifting out of resonance, not an eventual encounter with Mars, is
 responsible for depopulating the region. At the same time Morbidelli
 (1996) reached a somewhat different conclusion, though one still compatible 
 with the existence of a gap. His 3-dimensional integrations of 10 orbits 
 with $0.055 < e < 0.155$ indicated that all became Mars crossers in times 
 between 10 and 100~Myr. (His integrations were terminated once 
 this orbit crossing was noted.) The same fate happened to most orbits with 
 higher $e$'s, though four remained for a full $10^9$~years. Most objects 
 remaining at that time showed strong evidence that their proper
 eccentricities were diffusing toward higher values. 
 
 The more rapid crossing times found by Morbidelli may be the
 consequence of adding the third dimension to the problem. In a parallel
 development Henrard et al (1995) showed analytically that a strong 
 (nodal) secular resonance overlaps several secondary resonances. They
 suggested that it may be possible for a body initially in a low e secondary
 resonance to diffuse or random walk with an increasing libration amplitude
 so as to enter the extensive field of principally nodal secular resonances
 that lie at much larger $e$'s and $i$'s. Their short integration times
 of only 1~Myr failed to provide an example. At this point in
 our discussion the 1998 paper of Moons et al assumes special importance
 because it indicates that the chaotic areas arising from various sources
 are not isolated islands, but are connected. Bodies may therefore diffuse
 from low to high eccentricity and escape with only small changes in
 semimajor axis, much as was found in the numerical study just mentioned.
 How and when these three islands, located by
 Moons et al (1998) and
 containing very regular orbits, have lost bodies is, despite Morbidelli's
 long integrations, still an unsolved problem.  We would argue for additional
 very long term studies to quantify the process of diffusion especially for 
 orbits with small-to-moderate librations in the $0.2 < e < 0.45$ range 
 before placing all hope in an explanation linked to planetary migration 
 (cf Nesvorn\'y \& Ferraz-Mello 1997). Despite our present state of
 uncertainty, there is now a
 general conviction that strictly dynamical effects linked to Jupiter and 
 Saturn in their present orbits are still the best bet to exhaust a primordial
 population of bodies at the 2:1 resonance, but that such processes needed 
 more than a billion years to yield the current minor planet distribution.
 
 Figures~\ref{FigH}{\it a,b,c}, the companion to
 Figures~\ref{FigG}{\it a,b,c}, present parallel results for the 3:2
 mean-motion resonance, which is characterized not by a broad gap but
 by a  concentration of nearly 200 bodies (called the Hilda group) in reliable 
 (i.e. observed for two or more oppositions) librating orbits. The difference
 in the observed appearance at the two resonances is clearly reflected in
 their dynamical behavior. Although the strongest secondary resonance at 
 $P_{apse}/P_{lib} = 2/1$ is easily identified in Figure~\ref{FigH}, it is
 less deep (less chaotic) and far narrower than is the case at 2:1.
 Even when the eccentricities of Jupiter and Saturn are doubled
 (cf.Figure~\ref{FigH} {\it c}) its width is far less than its counterpart at
 2:1. A notable feature of the Hildas is the absence of bodies with
 proper eccentricities, $e_p \simlt 0.10$. Figure~\ref{FigH}{\it a} shows that chaotic 
 orbits dominate in the region $e_0 \simeq e_p < 0.05$ and that a mix of chaotic
 and regular ones lie from $e_0 = 0.05$ to the 2/1 secondary resonance at 
 $e_0 = 0.074$ with regular ones alone present at higher $e_0$'s, except 
 exactly at other secondary resonances. 
 
 Nesvorn\'y \& Ferraz-Mello (1997) have compared the long-term
 behavior throughout the 2:1 and 3:2 resonances. Their approach applies 
 the frequency map technique discussed by Laskar (1993) to determine the
 diffusion of (Fourier) frequency components in a resonant body's
 pericentric motion. The results show that all orbits in the heart of
 the 3:2 with $0.05 < e < 0.35$ diffuse very slowly, with percentage
 frequency shifts of about 10\% in a billion years. At 2:1 the broad, 
 continuous region found at 3:2 is replaced by numerous smaller islands also
 having slow diffusion, but they are surrounded by a sea of orbits where 
 diffusion is some 10 times faster. These plots are compatible with surveys 
 providing Lyapunov times, but they are quicker to obtain and hence more 
 complete. By either criterion, however, disruptive processes are one to
 two orders more severe and/or more effective over wider ranges of
 semimajor axis and eccentricity at 2:1. Efforts along these lines imply
 that the population difference between these two resonances is principally
 a matter of time scale: Wait long enough and the Hildas too will probably
 disappear, leaving behind a sort of undefined Kirkwood Gap at 3:2. We
 have integrated orbits of 10 hypothetical bodies with $0.02 < e_0 <
 0.075$ over the age of the solar system and found no signs of escape, but
 these integrations were only 2-dimensional. Longer integrations in three 
 dimensions are a useful topic to pursue. However, it is hard to rest content 
 with present and future population statistics alone. We would like
 understanding at a still more fundamental level: Why is the core of 3:2
 not broken into islands the way 2:1 is and, the related question, why are 
 Henrard's secondary resonances so much weaker there? Perhaps the latter 
 are a good starting point for future work. Henrard's studies have located 
 many of them as functions of eccentricity and inclination only within 
 the 2:1 resonance, but their relative and absolute strengths have yet to be 
 calculated. Despite much progress we do not yet understand the ultimate 
 reason for their curious behavior that is strong enough to produce marked 
 chaos at 2:1 over all $a$ and $e < 0.25$, but only in narrow slices at 3:2.
 
 The recent literature on the Kirkwood problem is extensive. For those 
 interested in more details, we recommend papers by Wisdom, Henrard and
 colleagues, Morbidelli, Moons and colleagues, Holman \& Murray
 (1996) and a
 review by the late Michele Moons (1997).

 \section{LONG-TERM STABILITY OF SMALL BODIES IN THE OUTER SOLAR SYSTEM}
 
 A vast number of asteroids occupy the region between Mars and
 Jupiter, and an even larger number of Kuiper belt objects exist
 near and beyond Neptune. Nevertheless, these objects contribute very
 little to the total mass of the solar system. The main asteroid belt
 and Trojan asteroids are estimated to contain a total of $\sim10^{24}-10^{25}$
 grams and the Kuiper belt no more than an Earth mass of
 material. In comparison to the mass of the planets themselves, this
 additional mass is almost negligible. In this sense, the regions
 between the planets are remarkably empty. Is this scarcity of
 material the result of particular processes of planet formation? That
 is, were those processes so efficient that nearly all of
 the initial mass was either incorporated into planets or swept away?
 Alternatively, are the gravitational perturbations of the present
 planets sufficient to eject nearly all of the material initially
 between the planets on time scales less than the age of the solar
 system? 
 
 Of course, these two ideas are not independent. The
 gravitational perturbations from proto-planets constitute
 one of the main physical processes during planet formation. The
 presence of proto-planets influences the orbital distribution of
 planetesimals that may or may not be accreted. Given the broad range of
 relevant physical processes and the computational challenge of including
 a sufficient number of bodies, direct simulations of planet formation
 are still severely limited. 
 
 It is much more straightforward to evaluate the gravitational
 influence of the present planetary system on smaller bodies.
 As we will see, the perturbations of the present day planets are
 sufficient to eject nearly all material from between the planets on
 time scales less than the age of the solar system. However, one cannot
 conclude that no other important physical process contributed to the
 absense of material between the planets. We will also see that
 there are regions in the solar system in which the time scale for
 removal by gravitational perturbations alone exceeds the age of the
 solar system. We will find dynamically long-lived regions that are
 empty of material, long-lived regions in which the orbital
 distribution appears excited by perturbations of bodies that are no
 longer present, and regions with dynamical life times that straddle
 the age of the solar system. It is in these regions that simulations
 of the long-term dynamics provide the richest evidence of the
 conditions of planet formation. In the following sections we discuss recent results on the long-term
 stability of small bodies in the regions between outer and inner
 planets.
 
 \subsection{The Region between Jupiter and Saturn}
 
 The abrupt decrease in the surface density of asteroids beyond the 2:1
 mean-motion (near 3.3~AU) revealed by the Palomar-Leiden Survey (van
 Houten et al 1970) prompted Lecar \& Franklin (1973) to study numerically
 the possibility that there were initially asteroids beyond where
 they are presently found. In addition to examining the long-term
 stability of small bodies in the outer asteroid belt, they examined the stability of such objects
 between Jupiter and Saturn. Modeling Jupiter and Saturn as moving on
 fixed elliptical orbits and considering the planar case, they
 integrated 100 test particles started on orbits between 5.7 and 9.1~AU
 with eccentricities between 0.0 and 0.1. After numerically
 integrating for 500 Jupiter periods ($\sim 6000$ years), only test
 particles near 6.8 and 7.5~AU remained.  The others escaped.
 Lecar \& Franklin (1973) cautiously concluded that the Jupiter-Saturn
 region would be depleted of asteroids in a few thousand years, with
 the possible exception of the two identified bands. Everhart (1973),
 in an independent study, identified the same long-lived bands in
 integrations lasting 3000 Jupiter periods. He then selected one test
 particle from each band and numerically integrated them until their
 orbits behaved chaotically (7,100 and 17,000 Jupiter periods,
 respectively).
 
 Franklin et al (1989) reexamined this problem, armed with
 faster computers that would permit longer numerical integrations.
 Although again the planar problem was considered, the effect of the
 mutual gravitation of the planets was modeled according to the
 leading two terms of the secular theory (see Murray \& Dermott 1999).
 The authors chose planet crossing orbits as their criterion for
 stopping a test particle integration. That is, if the test particle
 crossed the orbit of Jupiter or Saturn it was said to be ejected. Of
 the 135 test particle orbits integrated between Jupiter and Saturn,
 none survived. The longest lived was ejected after 799,000 Jupiter
 periods ($9.4\times 10^6$ years). The authors concluded that
 low-eccentricity, low-inclination orbits between Jupiter and Saturn
 were unlikely to survive longer than $10^7$~years. In addition to being
 the first to determine the full range of dynamical lifetimes of test
 particles between Jupiter and Saturn, the authors found that all 
 orbits displayed a positive Lyapunov exponent (were chaotic) before
 ejection.
 
 Soper et al (1990) extended this work to find a
 correlation between the estimated Lyapunov times and the ejection
 times of the test particles between Jupiter and Saturn. In addition,
 they established that their results were not sensitively dependent
 upon the numerical accuracy of the integrations. Even for
 dramatically degraded accuracy, stable orbits in the circular
 restricted three-body problem remain stable.
 
 Weibel et al (1990) also reexamined the problem of
 stability between Jupiter and Saturn. The principal improvements of 
 their work over previous studies were to integrate the actual orbits of Jupiter
 and Saturn and to integrate the full three-dimensional problem.
 Studying a sample of 125 test particles with initially
 low-eccentricity, low-inclination orbits, they found that nearly all
 were  planet crossing or ejected within 20,000 years, with a small number
 surviving more than $10^5$~years, in agreement with other results.
 Weibel et al (1990) also associated some of the variation in
 dynamical lifetime with the locations of mean-motion resonance with
 Jupiter or Saturn. 
 
 \subsection{The Regions Between the Other Outer Planets}
 
 As in the case of the outer asteroid belt and the region between
 Jupiter and Saturn, the question of whether there are regions in the
 outer planet region where small bodies might be stable on time scales
 of $10^9$ years has been raised numerous times in the
 literature. This question has been investigated by several groups
 using a variety of techniques. 
 
 Duncan et al (1989) developed an algebraic mapping to
 approximate the motion of a test particle orbiting between two
 planets. The mapping is composed of a part that follows the motion
 between conjuctions with a planet and a part that includes the
 impulsive gravitational influence at conjuction. Although the
 assumptions required to make its development tractable are severe, the
 mapping nevertheless recovered the size of the chaotic zone near a planet
 (Wisdom 1980) and the instability of test particles between Jupiter
 and Saturn. This mapping's speed relative to direct numerical
 integration permitted simulations lasting of order $10^9$~years, well
 beyond what could be completed at the time with available computers
 and conventional algorithms. With their mapping, the authors
 identified bands of long-term stability between Saturn and Uranus,
 Uranus and Neptune, and beyond Neptune. 
 
 Gladman \& Duncan (1990) were the first to complete accurate, direct
 numerical integrations of test particles in the Saturn-Uranus and
 Uranus-Neptune regions and beyond Neptune, as well as in the outer
 asteroid belt and Jupiter-Saturn region. Although their integrations
 were limited to 22.5~Myr by computational speed, they followed the
 trajectories of roughly one thousand test particles. Their additional
 advance was to include the perturbations of the four mutually
 interacting giant planets as perturbers, integrating individual test
 particles until they entered the gravitational sphere of influence of
 one of the planets. In addition to finding a clearing in the outer
 asteroid belt associated with mean-motion resonances and short
 time-scale dynamical erosion just beyond Neptune, they found that the 
 majority of test particles between the giant planets undergo close
 approaches with the planets in $10^5-10^7$ years. Whereas some test
 particles between each of the planets were found to survive, the
 authors noted that they did not expect them to be stable over the 
 lifetime of the solar system.
 
 Extending the work of Gladman \& Duncan (1990), Holman \& Wisdom
 (1993) studied test particle stability in the invariable plane from 5
 to 50~AU. Placing a total of 3000 test particles in circular orbits
 in the invariable plane (500 test particle in each of 6 initial
 longitudes), they integrated the particles for up to 800~Myr interior
 to Neptune and 200~Myr exterior to Neptune. This was subsequently
 extended to 4.5~Gyr interior to
Neptune and 1.0~Gyr exterior to Neptune by Holman (1995). The roughly order of magnitude speed-up in numerical
 integrations gained by the
 symplectic mapping method of Wisdom \& Holman (1991) allowed the
 more complete study. Duncan \& Quinn (1993), who
 approximated the motions of the outer planets by linear secular
 theory reported similar results. 
 
 Figure~\ref{Figtp} displays the dynamical lifetime
 as a function of initial semimajor axis. The solid line marks the
 minimum survival time of the six test particles initially in each
 semimajor axis bin. The encounter times of the test particles at
 the other initial longitudes for each semimajor axis bin are plotted as
 points, with surviving test particles plotted as open circles. 
 A number of dynamical features are immediately apparent. In
 each semimajor bin there is a fairly broad range of dynamical
 lifetimes, sometimes two orders of magnitude. As noted
 above, between Jupiter and Saturn nearly all test particles are
 removed by $10^5-10^6$~years. Test particles between Saturn and
 Uranus are nearly all removed by $10^8$~years, and those between
 Uranus and Neptune by $10^9$~years. Other than test particles librating 
 about the triangular Lagrange points of one of the planets and test
 particles beyond Neptune, only a single test particle survived.
 
 As Holman (1997) demonstrated, even this one surviving test particle
 does not represent a stable region between Uranus and Neptune.
 In the region 24--27~AU a small fraction (0.3 per cent) of a
 population of initially low eccentricity, low inclination orbits will
 survive 4.5~Gyr.  This regions is long-lived but not indefinitely
 stable.
 
 Recently, Grazier et al (1999a,b) revisited the issue of test particle
 stability in the Jupiter-Saturn, Saturn-Uranus, and Uranus-Neptune
 regions. They placed roughly 100,000 test particles
 in the Jupiter-Saturn zone and 10,000 test particles in each of the
 Saturn-Uranus and Uranus-Neptune zones. They employed a high-order linear
 multistep integrator with round-off error minimization and a small
 time step to accurately integrate their trajectories. Although their
 choice of initial orbital distributions makes direct comparisons
 difficult, their results largely confirm earlier ones and provide
 more detailed information of the time dependence of the removal of
 material as a function of orbital distribution. 
 
 \subsection{The Inner Solar System}
 
 Although test particle stability in the outer solar system has been
 thoroughly studied by a number of groups, few corresponding studies
 of the inner solar system have been conducted.
 Mikkola \& Innanen (1995) numerically integrated a few hundred test
 particles in the inner planet region (0.3--4.0~AU) for times up to
 3~Myr. They included all the planets as perturbers. In addition, 
 they estimated the Lyapunov times of each of the test particles by
 integrating the tangent equations of the Wisdom-Holman mapping (see
 Mikkola \& Innanen 1999.  Figure~\ref{mi95} shows these results. The test
 particle trajectories in the figure display a wide range of Lyapunov
 times, $10^2-10^6$~years. As Mikkola \& Innanen (1995) point out, the
 longest Lyapunov times (least chaotic trajectories) are found in
 the vicinity of the main asteroid belt. Outside of the main asteroid
 belt most of the test particles developed large enough eccentricity
 in the course of the integrations to become planet crossing. Whereas
 not all planet-crossers were ejected in 3~Myr, the authors suggested
 that longer integrations would clear many of the remaining such
 objects. However, the authors did identify two narrow regions, one
 between Venus and Earth and one just beyond Earth, where one might
 expect to find long-lived asteroid orbits with low eccentricity and
 inclination. 
 
 Evans \& Tabachnik (1999) integrated approximately one thousand test
 particles in the region 0.09--2.0~AU for times up to 100~Myr. They,
 like Mikkola \& Innanen (1995), included the nine planets as
 perturbers. In addition, Evans \& Tabachnik integrated five test
 particles at different initial longitudes in each semimajor axis bin
 to test the resulting range of dynamical lifetimes. Figure~\ref{et99}
 shows the results of their study. On time scales of 100~Myr a large
 fraction of the objects were removed; however, long-lived regions can
 be seen. Evans \& Tabachnik (1999) fit logarithmic and power-law
 decay profiles to populations, extrapolating the surviving population
 to 5~Gyr. They found two regions that could possibly harbor dynamically
 long-lived populations, 0.09--0.21~AU (interior to Mercury) and
 1.08--1.28~AU (between Earth and Mars). A few low-eccentricity and low-
inclination asteroids in the latter region can be found in current
 asteroid catalogs.
 
 \subsection{The Kuiper Belt}
 
 Arguing that the surface density of primordial material in the
 solar system should not end abruptly beyond the outer planets,
 Edgeworth (1943, 1949) and Kuiper (1951) independently suggested that
 a disk of material might be found beyond Neptune. Edgeworth (1943, 1949),
 furthermore, proposed that such a disk might serve as a reservoir of
 short-period comets. Decades later, numerical investigations (Fernandez 1980,
 Duncan et al 1988, Quinn et al 1990) showed
 that the orbital distribution of short-period comets is more
 consistent with an origin in a flattened, extended disk than in an
 isotropic distribution such as the Oort cloud (Oort 1950). 
 
 The discovery of the first  Kuiper belt object by Jewitt \& Luu
 in 1992 (Jewitt \& Luu 1993) and the subsequent discovery of nearly
 400 such objects has transformed the study of the trans-Neptunian
 region from a purely theoretical endeavor to one that is
 observationally grounded. For a recent review of the physical and
 observational aspects of the Kuiper belt we direct the reader to the
 recent chapter by Jewitt \& Luu (2000).
 
 The study of the long-term dynamics of the Kuiper belt is a rapidly
 maturing field with a rich literature. We describe
 only research that pertains to the issue of dynamical chaos. [For
 broader reviews of Kuiper belt dynamics see Morbidelli (1998) and
 Malhotra et al (2000). For a recent review of the formation and
 collisional evolution of the Kuiper belt see Farinella et al (2000)].
 
 In the first numerical experiments to examine the importance of
 dynamical chaos in the Kuiper belt, Torbett \& Smoluchowski (1990),
 improving upon the work of Torbett (1989), estimated the Lyapunov times of a
 large number of test particles with orbits beyond Neptune. Their
 10~Myr integrations included the four giant planets moving on
 fixed ellipses as perturbers. They identified a large chaotic zone
 that roughly coincides with test particle perihelia between 30 and
 45~AU. The Lyapunov times in this zone are less than
 300,000~years. Torbett \& Smoluchowski also noted that a small
 fraction of the test particles in this chaotic zone exhibit sizable
 diffusion throughout the zone. In addition, a fraction of the
 material in the belt could be scattered to large semimajor axis and
 effectively stored, forming a reservoir of comets. 
 
 Holman \& Wisdom (1993) and Levison \&
 Duncan (1993) directly demonstrated the viability of the Kuiper belt
 as a reservoir of short-period comets. Their numerical integrations
 showed a mixture of stable and unstable regions beyond Neptune.
 Small bodies in low eccentricity, low inclination orbits in some regions
 of the Kuiper belt can be delivered to Neptune-encounter orbits on time
 scales of
 $10^7-10^9$~years, with hints of instability on longer time scales (see
 Figure~\ref{Figtp}). Other regions appear stable for longer than
 $10^9$-year time scales. This is an essential point because an effective source of short-period
 comets must possess regions that are unstable on time
 scales comparable to the age of the solar system. Dynamical lifetimes
 significantly shorter than the age of the solar system would imply a now-depleted reservoir; a
 significantly longer dynamical time scale would imply an inadequate
 supply of short-period comets. Indeed,
 detailed calculation of the dynamical evolution of small bodies upon
 exiting the Kuiper belt or its extended component demonstrate that it
 is the likely source of short-period comets (Levison \& Duncan 1997, Duncan \& Levison
 1997).
 
 Duncan et al (1995) improved upon this early work by mapping
 the dynamical lifetimes in the Kuiper belt for a range of semimajor
 axes, eccentricities, and inclinations. Figure~\ref{FigDLB} displays
 the principal results of this study. The long-lived region can be
 described as those semimajor axes and eccentricities that give
 perihelia greater than 35~AU, with the exception of an unstable band
 between 40 and 42~AU associated with the overlap of secular resonances
 (Kne\v{z}evi\'c et al 1991, Morbidelli et al 1995).
 Although figure~\ref{FigDLB} reveals a rich dynamical structure, the
 underlying dynamics or causes of chaos and instability in the Kuiper
 belt were not explored in detail by Duncan et al (1995). 
 
 Two complementary approaches have been used to investigate the
 dynamical structure of the Kuiper belt. Morbidelli et al (1995)
 applied tools developed for the study of dynamics in the
 asteroid belt (Morbidelli \& Moons 1993, Moons \& Morbidelli 1995)
  to the Kuiper belt. Their approach was to use the planar circular
 restricted three-body problem, averaged for a particular resonance, to
 simplify the problem to a single degree of freedom. From that model the
 widths of a mean-motion resonance, in a semimajor axis, as a function
 of semimajor axis were computed. Morbidelli et al
 (1995) used similar
 models to examine the dynamics in secular resonances outside of
 mean-motion resonances. They pointed out that the unstable region
 40--42~AU at low eccentricity and the large eccentricity excursions seen 
 there by Holman \& Wisdom (1993) result from the interaction of the $\nu_8$
 and the $\nu_{18}$ secular resonances. Likewise, in the region
 35--36~AU large scale chaos results from the interaction of the $\nu_7$
 and $\nu_8$ secular resonances. The basic limitation of this
 approach, as the authors noted, is that each resonance must be examined
 in isolation to reduce the problem to a tractable single
 degree of freedom. Whereas these models can accurately describe the
 overall dynamics, the chaos that results from overlapping resonances
 is eliminated. 
 
 Malhotra (1996) used an alternative approach to map the boundaries of
 quasi-periodic regions associated with mean-motion resonances in the
 Kuiper belt. Surfaces of section of the circular restricted three-body
 problem show a divided phase space, with quasi-periodic regions
 interspersed with chaotic zones. Near a given mean-motion resonance
 the surfaces of section will show a stable island corresponding to the
 stable range of libration amplitudes or semimajor axis oscillation
 for a given eccentricity. Malhotra used the results from a series of
 surfaces of section to establish the stable boundaries of mean-motion
 resonances. These boundaries are somewhat narrower than those
 computed by Morbidelli et al (1995) because the analytic models can
 not account for the chaotic zones. Although the approach of using
 surfaces of section of the circular restricted three-body problem
 captures some the important effects of dynamical chaos at first-order
 mean-motion resonances, it also has a
 fundamental limitation. The eccentricity of Neptune's orbit must be
 ignored and only the planar case considered in order to reduce the problem to
 two degrees of freedom, from which a useful section can be computed.
 Thus, secular resonances from Neptune or other planets cannot be
 included. However, surfaces of section could be used to explore the
 dynamics in the regions of overlapping secular resonances in the Kuiper
 belt (see \v{S}idlichovsk\'y 1990). 
 
 Aside from establishing the general framework of stability in the
 Kuiper belt, the dynamical behavior in the 2:3 mean-motion resonance
 with Neptune has been studied extensively by a number of groups. This
 particular resonance has attracted a great deal of attention
 because, in addition to Pluto, a sizable population of Kuiper belt
 objects resides there. Whereas the orbit of Pluto and the resonances it
 occupies have been long established and well studied (see Malhotra \&
 Williams 1997, the range of orbital parameters of the known 
 Plutinos motivated broader studies of the 2:3 dynamics. Morbidelli
 (1997) examined the orbital diffusion throughout the resonance,
 finding dynamical lifetimes ranging from $10^6$~years to times in excess of the
 age of the solar system. In addition, Morbidelli
 (1997) studied the
 role played by the $\nu_8$ and $\nu_{18}$ secular resonance and the
 Kozai resonance within the 2:3 mean-motion resonance libration region.
 Related work on the dynamics in this resonance has been reported by
 Gallardo \& Ferraz-Mello (1998) and Yu \& Tremaine (1999) 
 The
 importance of dynamical scattering among different members of the 2:3
 resonance has also been recently examined by a number of groups (Ip
 \& Fernandez 1997, Yu \& Tremaine 1999, Nesvorn\'y et al 2000).
 
 Although other Kuiper belt mean-motion resonances have not been
 studied with as much detail as has the 2:3, the whole suite of
 resonances plays an important role in determining the overall structure
 and extent of the belt. The discovery of the first 
 scattered disk object, 1996~TL66, along with the recognition that the
 population of such objects must be substantial (Luu et al 1997),
 confirmed the suggestion of Torbett \& Smoluchowski (1990) that
 scattered Kuiper belt objects could be effectively stored at great
 heliocentric distances. Independent work by Duncan \& Levison (1997)
 at the time of this discovery immediately demonstrated by long-term
 numerical integration the mechanism of this storage. As a Kuiper belt
 object begins to undergo close approaches to Neptune, presumably after
 developing a large eccentricity in an unstable but long-lived region
 of the belt, the object's orbit follows a modified random walk.
 Successive encounters with Neptune alter the semimajor axis and
 eccentricity of the object's orbit in a way that roughly preserves
 perihelion distance (which is near Neptune). As a resonant value of
 the semimajor axis is approached, the random walk is altered. In some
 cases, as Duncan \& Levison (1997) demonstrate, temporary resonant
 trapping occurs, sometimes with a reduction in eccentricity that raises
 the perihelion distance beyond the immediately influence of Neptune.
 This effect was first discussed by Holman \& Wisdom (1993). These
 orbits, although trapped for very long times, will eventually develop
 large enough eccentricities to begin encountering Neptune again. By
 this means the scattered disk serves as an effective reservoir.
 It is clear that all of the trajectories that exhibit long-term
 capture in resonance are chaotic despite being long-lived. Although
 numerical integrations have demonstrated this, there is little
 analytic work on the details of this capture. Such work would provide
 valuable insight into how material in the extended Kuiper belt is
 distributed. 
 
 \section{\bf PLANETARY CHAOS}
 
 \subsection{Numerical Integrations}
 
 By the 1980s it was clear that most Hamiltonian systems exhibited both
 chaotic and regular (on tori) motion. The chaotic motion is intimately
 tangled up with regular motion on KAM tori. However, the prevailing
 feeling was that the solar system was almost certain to lie on a KAM
 torus.
 
 This expectation seemed to be suported by early attempts at accurate
 long term integrations of the solar system, including those of
 Applegate et al (1986), who carried out integrations over 3~Myr
, neglecting Mercury. The LONGSTOP project integrated the outer solar
 system (Jupiter through Pluto) using a standard general purpose
 integrator for a time of 9.3~Myr (Milani et al 1986). In these and other
 integrations the planets did nothing untoward. Applegate et al also
 carried out 200~Myr integrations of the outer planets. The motion
 appeared to be multiply periodic, as expected of motion on KAM tori,
 although they noted the presence of very long period variations in
 Pluto's orbital elements.
 
 It was therefore a surprise when Sussman \& Wisdom (1988) showed
 that the orbit of Pluto was chaotic. They used a special purpose-built
 computer called the Digital Orrery, running a twelfth-order Stormer
 integrator. This work featured the first attempted measurement of the
 Lyapunov exponent of the planetary system. The Lyapunov exponent is a
 standard tool in the arsenal of nonlinear dynamics, designed
 specifically to see if a system is chaotic. If the separation grows
 exponentially with time, $d(t)\sim e^{t/T_l}$, the orbit is chaotic. 
 Multiply periodic orbits lead to much slower power law separation with
 time, $d(t)\sim t^\alpha$.
 Sussman \& Wisdom (1988) found that the orbit of Pluto was chaotic,
 with a Lyapunov time of $T_l\sim 20$~Myr.  Later, the LONGSTOP integrations
 were extended (Nobili et al 1989). This paper
 did not examine Lyapunov times, but it suggested, based on the
 appearence of the Fourier spectrum, that the orbits of the outer
 planets might be chaotic.
 
 At roughly the same time Laskar (1989) performed numerical
 integrations of a very different type of model. He solved a subset of
 Lagrange's equations for the orbital elements; Lagrange's equations
 are similar to Equation (\ref{lagrange}).
 Laskar's model consisted of
 analytically averaged equations describing the motion of all the
 planets except Pluto. In this model he kept secular terms up to second
 order in the planetary masses and to fifth order in eccentricities
 and inclinations (Laskar 1985).  He also included the analytically averaged secular effects of all mean
 motion terms up to the same order. This involves dropping any term
 exhibiting a sinusoidal function whose argument contains a mean longitude.
 However, it does account for the (secular) effects of terms proportional
 to the product of two such sinusoids. For example, consider
 Equations (\ref{lagrange}), (\ref{expansion}), and (\ref {a_52}). Every
 term in the disturbing function is proportional to $M_2$. To lowest order
 in mass the simple averaging procedure employed consists of
 dropping every term in (\ref{expansion}) that contains a mean
 longitude in the argument of cosine.
 
 However, Laskar (1989) considered terms of second order in the
 masses. For example, consider using Equation (\ref{mean_motion}) in 
 an extended development of (\ref{lagrange}).
 The right hand side of the latter will contain a term proportional to
 \ba 
 4a_1{a_1\over a_2}{M_2\over M}
 {n_1\over  2n_1-5n_2+3\dot\varpi_1}\phi_{2,-5,3,0,0,0}(a_1,a_2)e_1^3
 \cos[2\lambda_1-5\lambda_2+3\varpi_1]\nonumber\\ 
 \times -2{{\cal  G}M_2\over a_2}\phi_{2,-5,2,1,0,0}(a_1,a_2)e_1^2e_2
 \sin[2\lambda_1-5\lambda_2+2\varpi_1+\varpi_2].
 \ea 
 Using the  trigometric identity
 $\cos(a+b)\sin(a+c)=(1/2)\sin(2a+b+c)+(1/2)\sin(c-b)$, we see that
 this term will give rise to a factor $\sin(\varpi_2-\varpi_1)$. Because
 neither $\lambda_1$ or $\lambda_2$ appear, this secular term contributes
 to the
 averaged Hamiltonian. Terms that are nearly resonant, i.e. terms in which the combination
 $pn_1-qn_2$ are small, will produce relatively large (compared with the
 simple estimate $M_2^2$) contributions to the averaged Hamiltonian.
 Laskar's (1989) model contained some 150,000 secular and averaged terms.
 
 Laskar (1989) found by numerical integration over 200~Myr that in his model the
 entire solar system was chaotic, with $T_l\approx5$~Myr. He stated
 without explanation that ``the chaotic behaviour of the Solar System
 comes mainly from the secular resonances among the inner planets.''
 
 In a later paper Laskar (1990) showed that two combinations
 of secular angles appeared to alternate between libration and
 rotation, implying that the orbit crossed the separatrix of these
 resonances. Such behavior is associated with chaotic motion; in
 certain cases it may be the origin of the chaos. However, the two
 resonances Laskar identified involved the angles
 $\sigma_1\equiv(\varpi^0_1-\varpi^0_5)-(\Omega^0_1-\Omega^0_2)$ and
 $\sigma_2\equiv2(\varpi^0_4-\varpi^0_3)-(\Omega^0_4-\Omega^0_3)$. 
 The angle $\varpi^0_4$ is associated with the fourth normal mode
 of the planetary eccentricities, with a similar interpretation for the
 other $\varpi_0$'s. 
In some instances (such as Jupiter)
 $\varpi_5\approx\varpi_J$, but the $\varpi$'s are a combination of all 
the normal modes. Similarly the angle $\Omega^0_4$ is the angle
 associated with the fourth normal mode of the planetary
 inclinations. Because the two resonances identified by Laskar do not
 interact directly, they are unlikely to produce any substantial chaos.
 
 Because Laskar employed an averaged system of equations, it was
 important to verify his results using an unaveraged system of
 equations. This was done by Laskar et al (1992). They examined the numerical solution of Quinn
  et al, a 6~Myr integration of the entire solar system. Although this
 integation was not long enough to detect the chaos, it did allow them
 to verify that the resonant argument $2(\varpi_4^0 - \varpi^0_3) -
 (\Omega^0_4 - \Omega_3^0)$ alternately librated and rotated.
 
 Two years later Laskar (1994) identified a second secular
 resonance involving Earth and Mars;
 $\sigma_3\equiv(\varpi^0_4-\varpi^0_3)-(\Omega^0_4-\Omega^0_3)$. He
 noted that on some occasions $\sigma_2$ librated when $\sigma_3$
 rotated, and vice versa. This led him to suggest that the overlap of
 these two resonances was responsible for the chaotic motion.
 
 The next advance was the work of Sussman \& Wisdom (1992). They
 employed the Wisdom-Holman symplectic mapping to perform a 100~Myr
 integration of the entire solar system, which they found to be chaotic
 with $T_l\approx5$~Myr. This type of integration accounts for all
 types of resonance, both secular and mean-motion. They confirmed that
 the first two resonances identified by Laskar (1989) do exhibit both
 libration and rotation, but the second Earth-Mars, $\sigma_3$ resonance
 never
 librated, but only rotated, in their integrations. They were careful
 to point out that this did not rule out the interpretation of Laskar
 that the two resonances involving Earth and Mars overlap to cause the
 chaos.
 
 They also found two other combinations of angles that, in their
 integrations, both librate and rotate, namely
 $\sigma_4\equiv3(\varpi_4^0-\varpi_3^0)-2(\Omega_4^0-\Omega_3^0)$ and
 $\sigma_5\equiv (\varpi_1^0-\varpi_8^0)+(\Omega_1^0-\Omega_8^0)$. They
 found that four of the five angles (all but $\sigma_3$ showed a
 transition from libration to rotation, or vice versa, at roughly the
 same time). This strongly suggests to us that some, as yet unidentified,
 mechanism is forcing the transitions seen in the integrations. This
 point is reinforced by the observation that $\sigma_1$ and $\sigma_2$
 do not strongly interact.
 
 In addition to confirming Laskar's basic result that the entire solar
 system is chaotic, Sussman \& Wisdom found that the outer planets by
 themselves were chaotic, with $T_l\approx 7$~Myr. The Lyapunov times
 found in their giant planet integrations seemed to depend on the step
 size, at first glance a rather
 disturbing finding. However, a second set of integrations using a general
 purpose Stormer scheme again showed that the system was chaotic, this time
 with $T_l\approx19$~Myr.
 
 As a check that some long-term integrations of a planetary system were
 not chaotic, they carried out a 250~Myr integration of the outer
 planets without Uranus and found no evidence of chaos.
 
 These numerical experiments indicated that the solar system was
 chaotic, but there was no
 indication in any of the integrations that any of the planets would
 suffer either ejection from the solar system or collision with another
 body. In this sense it appeared that the solar system was stable.
 This comforting interpretation was bolstered by a 25~Gyr integration
 carried out by Laskar (1994). He found that none of
 the planets (excluding Pluto, which was not integrated) suffered an
 ejection or collision over that time. This suggested that the solar
 system was stable for $10^{10}$ years or more.
 
 Laskar's integration showed that the eccentricity of Mercury varied
 between $0.1$ and $0.5$, with an average value of about $0.2$. He
 attributed the variations to a diffusive process, driven by 
 chaos. If we assume that this is the case, we can estimate the
 diffusion coefficient, and hence the time to remove Mercury by
 collision with the sun or with Venus, when $e\to1$. The diffusion
 coefficient is
 \be 
 D\approx(\bar G-\bar G_0)^2/T,
 \ee 
 where $\bar G\approx e^2/2$ and $T$ is the length of the integration,
 25~Gyr. We estimate the maximum excursion in $\bar G$ using $e_0=0.2$
 and $e=0.5$, corresponding to $\bar G_0=2\times10^{-2}$ and $\bar
 G=0.125$. We make the assumption that the diffusion coefficent is
 independent of $e$, which is incorrect but adequate for our purposes.
 We find $D\approx 1.5\times10^{-2}/T$. The time for $e$ to diffuse to
 $1$ is
 \be \label{Mercury_escape} 
 \tau_{esc}\approx 1/D=T/1.5\times10^{-2},
 \ee 
 or about $2\times10^3$~Gyr, or $2\times10^{13}$ years. 
 Laskar then repeated the integrations several times, each time
 changing the eccentricity of Earth by about one part in a billion. The
 integrations differed in detail, but no collisions or ejections were
 seen. 
 
 One might question whether Mercury could actually diffuse to such a
 large eccentricity; might there be some dynamical barrier preventing
 it from doing so? In a series of numerical experiments Laskar (1994)
 showed
 that there were indeed orbits in his averaged equations that were very
 close to those of the solar system, and for which Mercury suffered a
 collision with the sun. In his words, he ``decided to guide Mercury to
 the exit.'' He made four clones of the Earth's averaged orbit, again
 changing the eccentricity by different amounts of the order of a part
 in a billion. He then integrated for 500~Myr. He retained the solution
 having the largest value of $e$ for Mercury, using it to produce four
 more clones with altered orbits for Earth. 
 
 Repeating this process 18 times, Laskar found a system in which the
 pseudo-Mercury was ejected after a 6~Gyr integration. This is much
 shorter than our estimate above, but this is to be expected because
 Laskar was actively searching for the most unstable orbit. The
 significance of the experiment is not that it predicts loss of Mercury
 on times comparable to the age of the solar system; the earlier
 experiments had already shown that the time for this to occur was
 longer than 25~Gyr. Rather, the experiment showed that it was
 plausible that there were no dynamical barriers to the loss of Mercury
 due to chaotic perturbations.
 
 Murray \& Holman (1999) carried out roughly 1000 long-term
 integrations of the outer solar system using the Wisdom-Holman
 symplectic mapping. They investigated the effect of altering the
 semimajor axis $a_U$ of Uranus, tracing out the variation of $T_l$ as
 a function of $a_U$. Using this technique, they located chaotic regions
 associated with the 2:1 resonance between Uranus and Neptune, the 7:1
 resonance between Jupiter and Uranus, and with three-body resonances
 involving Jupiter, Saturn, and Uranus and Saturn, Uranus, and Neptune
 (see the section on analytic results, below). The variation of $T_l$
 with $a_U$ is shown in figures~\ref{Fig_TLa} and \ref{Fig_TLb}. The few hundred million year length
 of most of the runs limited their ability to place lower limits on
 $T_l$ to about $100$~Myr. More recently, we have extended some runs to
 $1$~Gyr; the results of two runs are shown in Figure
 \ref{Fig_long}. The figure plots the phase space distance between two
 copies of the solar system, in which one copy of Uranus is displaced
 relative to the other by about $1$ mm. There are two such calculations
 displayed, corresponding to two different fiducial values of $a_U$,
 $19.23$, and $19.26$. One is chaotic, the other is regular, or has
 $T_l$ larger than about 0.5~Gyr. Note that an integration of less than
 200~Myr would indicate that both systems were regular.
 
 Murray \& Holman also showed that many planar four-planet models were
 chaotic, indicating that inclination resonances were not required to
 produce chaotic motion in the outer planets. They demonstrated that a
 three-planet, nonplanar system without Neptune was often chaotic.
However, a three-
planet system with no Uranus or a three-planet planar problem with no
 Neptune was found to be completely regular, independent of the
 locations of the other planets (within moderate limits).
  one of the orbits, even in strongly chaotic systems, showed any sign
 of substantial changes in $a$, $e$, or $i$ over the length of the
 integrations. 
 
 This result was extended by Ito \& Tanikawa (2000) to the entire
 solar system, over several $\sim4$~Gyr integrations, using a
 Wisdom-Holman integrator. The integrations confirm the finding that the solar
 system is chaotic.  They also show that the orbits of the planets do not change appreciably
 over the age of the solar system; the full system of equations is not
 appreciably more unstable than Laskar's averaged system. An estimate of
 the diffusion time, similar to that
 given above, but using the results of Ito \& Tanikawa (2000),
 predicts that Mercury will not suffer any catastrophic encounters for $10^{3}$ or
 even $10^{4}$~Gyr. We appear to be safe for now.
 
 \subsection{Analytic Results}
 
 The numerical results described above suggest that the solar system is
 chaotic, with a Lyapunov time of about 5~Myr. This result is
 surprising, because the solar system is observed to be more than 4~Gyr
 old. Not so surprising is the result that the integrations are stable,
 in the sense that no close encounters, defined by one body entering
 the Hill sphere of another, are found. In fact, the results of Ito \&
 Tanikawa (2000) indicate that no planet has suffered even moderate
 changes in semimajor axis, eccentricity, or inclination. 
 Why does the solar system
 appear to be so chaotic and, if it is, why is it so resistant to
 catastrophe?
 
 Recently, we found an analytic explanation of both results, short term
 chaos and long term stability, in the setting of the outer solar
 system (Murray and Holman 1999). We start with the observation that chaos results
 from the interaction of at least two resonances between motion in two
 or more different degrees of freedom. We have to find the resonances.
 For example, consider the ``great inequality,'' the near 5:2 resonance
 between Jupiter and Saturn. The orbital period of Jupiter is
 4,332.588 days, whereas that of Saturn is 10,759.278 days, giving a
 ratio of 2.4833. The mutual perturbations of these planets produce
 large (relative to the mass ratio $M_J/M_\odot$ or $M_S/M_\odot$)
 variations in $\lambda$ when compared with the Keplerian value. The
 variation amounts to about 21 and 49 arcminutes in the longitude of
 Jupiter and Saturn, respectively, variations that were noted by
 astronomers in the eighteenth century. Mathematically, the resonance is
 represented by the following terms in the disturbing function:
 \be \label{great_inequality}
 -{{\cal G}M_S\over a_S}
 \sum_{k,q,p,r}\phi_{k,q,p,r}^{(2,5)}(a_S/a_J)e_S^ke_J^qi_S^pi_J^r\cos
 \left[2\lambda_J-5\lambda_S+k\varpi_S+q\varpi_J+p\Omega_S+r\Omega_J\right].
 \ee 
 Recall that $2-5+k+q+p+r=0$, and that $p+r$ must be even. Furthermore,
 $(2n_J-5n_S)/n_J\approx-1.33\times10^{-2}$; although this is small, it is much larger than
 the ratio of any of the secular frequencies with $n_J$; for example,
 $\varpi_S/n_J\approx2.6\times10^{-4}$. The small magnitude of the
 secular frequencies implies that including the secular
 frequencies will change the location (in semimajor axis) of the
 resonance only slightly. On the other hand, the rather large distance
 from exact resonance ($\sim10^{-2}$ is large compared with the width of the
 resonance, as we show below) shows that the planets are not ``in''
 resonance, i.e. none of the angles in the argument of the cosine in
 Equation (\ref{great_inequality}) librate.
 
 The last statement can be generalized: The only planets in the
 solar system involved in a two-body mean-motion resonance are Neptune
 and Pluto. These two bodies are in a 3:2 mean-motion resonance, as
 well as a number of secular resonances. The chaos seen in
 integrations of the giant planets, and in the solar system excluding
 Pluto, is not due to the interaction of two-body mean-motion resonances.
 
 The fact that Jupiter and Saturn are not in resonance does not mean
 that the resonant terms given by (\ref{great_inequality}) are
 negligible, however. They produce substantial variations in
 the semimajor axis (given by Equation \ref{a_52} and similar terms)
 and in the longitudes of the two planets; it was the latter, which involve
 two powers of $(2n_J-5n_S)/n_J\approx-1.33\times10^{-2}$ in the
 denominator, that is responsible for the 21 arcminute discrepancy seen
 in the longitude of Jupiter by the eighteenth century observers. These
 near-resonant terms
 also produce substantial variations in the eccentricity and
 inclination of both planets.

For example, Saturn's gravity forces variations in $e_J\sin\varpi_J$ given by
 \begin{eqnarray} \label{ej} 
 e_J^{(2,5)}\sin\varpi_J&\approx& {\mu_S\over(2-5n_S/n_J)}{a_J\over a_S}\sum_{p>0} 
 \phi_{k,p,q,r}^{(2,5)} e_S^k e_J^{p-1}i_J^q i_S^r\nonumber \\
 & & \times\sin[2\lambda_J-5\lambda_S+k\varpi_S+(p-1)\varpi_J+q\Omega_J+r\Omega_S],
 \end{eqnarray} 
 where $\mu_S\equiv M_S/M$ is the mass ratio of Saturn to the Sun.
 The largest variation in $e_J$, corresponding to $k=2$, $p-1=q=r=0$
 and $\phi_{2,1,0,0}\approx 9.6$, has an amplitude of about
 $3.5\times10^{-4}$. Numerical integrations yield
 $3.7\times10^{-4}$ (Murray and Holman 1999). As shown in the following section, this variation in
$e_J$ plays a central role in
 producing chaos among the outer planets.
 
 \subsubsection{THREE-BODY RESONANCES}
 Although there are no two-body resonances between the giant planets,
 there are a number of resonances involving three bodies. Three-body
 resonances involve the longitudes of three planets; the combinations
 $3\lambda_J-5\lambda_S-7\lambda_U$ and
 $3\lambda_S-5\lambda_U-7\lambda_N$ are two examples. There are no
 terms containing such arguments in the disturbing function; they arise
 only at second order in the planetary masses. Physically, they arise as
 follows. 
 
 Consider Jupiter, Saturn, and Uranus. In the first
 approximation all three follow Keplerian orbits, so $a$, $e$, and $i$
 (as well as $\varpi$ and $\Omega$) are constant for all three
 bodies. At the next level of approximation, Saturn perturbs the orbit
 of Jupiter, and vice versa. This will, for example, cause tiny
 variations in $e_J$ and $e_S$, as calculated in the previous
 section. The amplitude of the variation will be proportional to the
 mass of the perturbing planet.
 
 Now consider the potential experienced by Uranus. To lowest order Uranus
 moves on a Keplerian orbit, so to first order in the masses it will
 see the potential given by the disturbing function with the Keplerian
 values of $e_J$ and so forth. At second order in the masses, several
 types of correction arise. One type is due to the fact that Uranus'
 orbit is not Keplerian. For example, Jupiter will force changes
 in $a_U$, $e_U$, and so forth, which have magnitude proportional to
 $M_J$ and period given by $pn_J-qn_U$, where $p$ and $q$ are
 integers. The position vector $\br_U$ will inherit oscillatory terms of
 this form. The potential experience by Uranus, due to Saturn, will in
 turn inherit terms proportional to $M_J$, with resonant arguments
 involving $\lambda_J$. This will lead to terms of the form
 $M_J M_S\cos[p\lambda_J-r\lambda_S-(s+q)\lambda_U]$. We refer to such
 terms as three-body resonances.
 
 Three-body resonant terms arise in two other ways. We have already
 said that Saturn will produce variations in Jupiter's orbital elements
 of the form $M_S\cos[l\lambda_J-m\lambda_S]$. The potential
 experienced by Uranus, assumed to be on a Keplerian orbit, contains
 terms of the form
 $M_Je_J^l\cos[p\lambda_J-q\lambda_U+l\varpi_J]$. 
 However,
 $e_J$ is no longer constant; $e_J(t)$ contains terms of the form
 $M_S\cos[p\lambda_J-r\lambda_S]$. Once again, these will give rise to
 terms proportional to $M_J M_S$ contaning resonant arguments involving
 all three mean longitudes.
 
 Similarly, Jupiter will produce variations in Saturn's orbital
 elements, which will in turn affect the potential experienced by
 Uranus and give rise to terms proportional to the mass of both
 Jupiter and Saturn, and having resonant arguments involving all three
 planetary mean longitudes.
 
 Murray and Holman (1997) gave analytic estimates of the strength, or width,
 and of the separation of such resonances. The width of a typical component resonance is
 \be 
 {\Delta a\over a_U}=8
 \sqrt{(6-p)\phi^{(7,1)}_{6-p,p,0,0}\phi^{(2,5)}_{2,1,0,0}{\alpha\over3\epsilon}
 \mu_J\mu_Se_J^{5-p}e_U^pe_S^2}\approx2\times
 10^{-6},
 \ee 
 where $\alpha = a_J/a_S \approx 0.55$ and $\epsilon = |2 - 5
 (n_S/n_J)|$ ($n_S$ and $n_J$ being the respective mean-motions of Saturn and
 Jupiter).
 This yields $\Delta a\approx8\times10^{-5}$ AU. The libration period
 associated with a resonance of this amplitude is 
 \be 
 T_0=T_U\Bigg/\sqrt{147(6-p)\phi^{(7,1)}_{6-p,p,0,0}\phi^{(2,5)}_{2,1,0,0}
 {\alpha\over\epsilon}\mu_J\mu_Se_J^{5-p}e_U^pe_S^2}\approx 10^7
 {\rm years},
 \ee 
 where $T_U$ is the orbital period of Uranus.
 This is essentially the Lyapunov time (Holman \& Murray 1996, Murray
 \& Holman 1997).

 Murray \& Holman (1999) estimate the time for Uranus to suffer a close
 encounter with Saturn. An ejection or collision would then follow in
 short order. The estimate assumes that there are no dynamical barriers
 to the random walk of $e_U$ produced by the chaos. They find a time of
 order $10^{18}$ years, much longer than the current age of the
 Universe.
 
 Figures~\ref{Fig_TLa} and \ref{Fig_TLb} show the location of various two- and three-body
 resonances in the vicinity of Uranus. In figure~14 one can
 see individual three-body mean-motion resonaces. The resonant argument
 of the resonance closest to the best estimate of the orbit of Uranus
 is seen to alternate between libration and rotation in figure (\ref{Fig_jsu})
 
 \subsubsection{CHAOS IN THE INNER SOLAR SYSTEM}
 The situation in the inner solar system is currently unclear. There
 have been a number of candidate resonances suggested, but no analytic
 calculations have been done. This is clearly an opportunity for an
 enterprising theorist. 
 
 Without a calculation in hand one cannot say
 what Lyapunov time one expects from overlap of secular resonances, or
 how to predict the time required for a planet's (Mercury in this
case) orbit to
 change drastically. We can use estimates similar to that given in
 Equation (\ref{Mercury_escape}), but they are on rather shaky ground
 because we do not know if the variations of $e$ we see in the
 integrations are primarily diffusive or if they are actually the
 result of quasiperiod forcing of Mercury's orbit by, e.g. Venus and
 Earth. There is some evidence for the latter, because both Laskar
 (1990, 1994), Laskar et al (1992), Sussman \& Wisdom (1992),  and Ito \& Tanikawa (2000) find
 strong correlations between the motion of all three planets. If the
 variations in the eccentricity of Mercury are primarily due to
 quasiperiodic forcing, then Mercury's lifetime could be much longer than our estimate.
 
 From the numerical results of Laskar (1994) and Ito \& Tanikawa (2000) we gave a
 rough estimate of $10^{13}$ years for the lifetime of Mercury. Murray
 \& Holman (1999) found $10^{18}$ years for the lifetime of Uranus. In units
 of orbital periods these lifetimes are $4\times 10^{13}$ and
 $10^{16}$, a ratio of about 250, yet the Lyapunov times of the two
 systems are within a factor of about two. Without an analytic theory for
 the chaos in the inner solar system it is difficult to assess the
 significance of this discrepancy.
 
 We have noted that the resonance $\sigma_1$ identified by Laskar does
 not overlap with any other secular resonance that has so far been
 identified. This suggests that it is not the source of the chaotic
 motion seen in various integrations. Rather, it appears that the
 transitions between libration and rotation are the result of chaotic
 forcing by other planets. This may be checked by integrations of the
 solar system excluding Mercury, in which Uranus is moved to a location
 outside the chaotic three-body mean-motion resonances. If the
 resulting system is still chaotic, then the resonance corresponding to
 $\sigma_1$ does not play an essential role in producing the chaos seen
 in the integrations.

 \section{SUMMARY AND SUGGESTIONS FOR FURTHER WORK}
 
 The solar system is unstable, although on times much longer than
the Lyapunov times. Our main task is to identify the resonances
that overlap and induce chaos and to predict the ejection time
as a function of the Lyapunov time (which is relatively easy to
calculate). For example, in the region of overlapping first-order mean-motion
resonances the ejection time is proportional to the Lyapunov time
to the 1.75 power. A similar power law relation, with a different
exponent, holds for high-order mean-motion resonances, where the various
subresonances overlap. As yet, we have no such relation for secondary
resonances, where the libration frequency is a multiple of the apsidal
motion.
Surprisingly, the relation with the exponent of $1.75$ holds approximately throughout the solar
system, although the spread in ejection times for 90\% of the trajectories
is a factor of 10 on either side of the prediction, (see figure~\ref{Fig_tl_te}). The
exceptions occur
at high-order mean-motion resonances or at overlapping secular resonances.
The ``diffusion'' among the subresonances inside the same mean-motion
resonance is much slower than the diffusion between overlapping first-order
mean-motion resonances.
The relevant model for chaos in the solar system is the overlap
 of two resonances; in the Hamiltonian formulation this resembles
 a pendulum driven at resonance. This induces a random walk
(diffusion) in the
 eccentricity that can result in a close encounter with the 
 perturber and a radical change in the orbit. We do not believe that
 a web of resonances (the Arnold Web) is relevant for chaos in the
 solar system, as interesting as that formulation is mathematically,
 and we no longer believe that instability is caused by a secular
drift in the semimajor axes.
 
Three-body resonances have been identified as the source of chaos for
the outer planets. Because these resonances are proportional to the
product of the mass ratios of the two planets to the Sun, the time scale is
quite long, on the order of $10^9$ times the age of the solar system.

The identification of the overlapping secular resonances in the inner solar
system (the terrestrial planets) is not firm, but an extrapolation of
the numerical integration indicates that Mercury will be in trouble
in $10^{13}$ years (well after the Sun becomes a red giant and
engulfs Mercury). 

The work we have reviewed here could be termed `weak chaos'.  We
 picked up the story after the violent encounters associated with the
 formation of the Solar System were over.  The trajectories we studied
 could be treated by the well-developed methods of modern non-linear
 dynamics and celestial mechanics.  The discovery of extra-solar planets
 draws our attention to the era of formation when the planetary bodies
 were less well-behaved;  when close encounters, collisions, mergers,
 and ejections were the norm.  That was the era of `strong chaos'.
 Exploring this should provide the palette of stable configurations.

 \clearpage
 
 \begin{figure}
 \caption{
 ({\it a}) A case of (temporary) secular resonance within the 3:1 mean-
motion resonance. Dark solid line here and elsewhere marks the motion
 of Jupiter's apse, $\varpi_J$, showing the effect of the longer
 term $\nu_5$ and shorter $\nu_6$ variation. Crosses correspond to a body with
 $a_0 = 0.481$ [$a_{Jup} = 1.0$] and $e_0 = 0.05$.
 ({\it b}) Eccentricity, $e$, surges that develop from the case of secular
 resonance shown in ({\it a}). $e$'s $> 0.32$ cause a crossing of Mars' orbit. 
 Note that the condition $\varpi_A > \varpi_J$ corresponds to an increase in $e$.
 \label{FigA}}
 \end{figure}
 
 \begin{figure}
 \caption{
 Secular resonances within the 5:2 resonance, after Moons \&
 Morbidelli (1995). Two broad lines mark the limits (``separatrices'') of
 5:2, and the central line periodic solutions of the restricted three-body
 problem. Lower (upper) thin solid lines are the loci of the $\nu_6$ 
 ($\nu_5$) secular terms and the dashed lines, their approximate limits. Central 
 hatched region contains nonchaotic orbits (cf figure~\ref{FigD}d,e). Note that
 orbits with $e < 0.2$ have a vanishing chance of escaping the effects of
 both $\nu_5$ and $\nu_6$ and hence are especially chaotic.
 \label{FigB}}
 \end{figure}
 
 \begin{figure}
 \caption{
 Distribution of 58,000 asteroids with reliable orbits from
 the most current files of the Minor Planet Center. Principal
 mean-motion resonances are indicated. The gap/boundary near $a =
 2.08$~AU results from the strong perturbations on e's and i's due to the $\nu_6$
 and $\nu_{16}$ secular resonances. 
 \label{FigC}}
 \end{figure}
 
 \begin{figure}
 \caption{
 A typical example of a regular orbit, $a_0 = 0.481$,
 $e_0 = 0.40$, libration amplitude $8^\circ$ degrees, lying with the 3:1
 resonance. ({\it a}) shows that it is unaffected by secular resonance; ({\it b}) plots the
 separation in longitude with time for two bodies with the above 
 elements that are identical except for an initial longitude
 difference of $10^{-6^\circ}$. This body's essentially zero slope implies a
 very long Lyapunov time and consequently an orbit that shows no sign
 of being chaotic. ({\it c}) shows that its eccentricity variations guarantee
 frequent crossing of Mars' orbit, which requires only $e > 0.35$.
 \label{FigD}}
 \end{figure}

 \begin{figure}
 \caption{
 Examples of chaotic and regular behavior at the 5:2
 mean-motion resonance. The large amplitude oscillations of 
 $\varpi_A - \varpi_J$ in ({\it a}) lead to a chaotic orbit with a Lyapunov time, given
 by the slope in ({\it b}) of $\log T_l = 3.01$ in $P_J$. Its eccentricity
 regularly exceeds 0.5. The smaller oscillations shown in ({\it c})
 corresponds to a far more regular orbit with $log T_l > 5.6$. Despite
 formally lying in the $\nu_5$ secular resonance, such orbits show no signs
 of escape or $e$ increase beyond 0.35 in integrations of a planar model
 extending to 2 billion years.
 \label{FigE}}
 \end{figure}
 
 \begin{figure}
 \caption {
 Survival and escape times at resonances in the outer asteroid belt after
 Murray \& Holman (1997). Open and filled symbols correspond to predicted
 and numerical estimates, arrows to lower limits.
 \label{FigF}}
 \end {figure}
 
 \begin{figure}
 \caption{
 Locations of secondary resonances, defined by the ratio of apsidal to
 libration periods, lying within the 2:1 mean-motion resonance for a
 semimajor axis $a_0 = 0.630$ ($a_J = 1.0$). Vertical scale measures degree
 of chaos by plotting the Lyapunov time, $T_l$, in Jovian orbital periods.
 As the eccentricities of Jupiter and Saturn are artificially reduced in
 ({\it b}) and ({\it c}), the influence of secondary resonances falls and orbits
 become more regular. Note in the insert in ({\it a}) that a few regular orbits
 do exist exactly at the 3/2 secondary resonance even when the
 eccentricities of the two planets are not lowered. As indicated, they are
 also found at 7/1, 11/1, and 12/1. Proper eccentricities of hypothetical
 bodies are about 0.03 larger than the plotted initial values. Libration
 amplitudes range from $2^\circ$ at $e_0 = 0.25$ to $60^\circ$ at $e_0 = 0.02$. All
 orbits plotted at $log T_l > 5$ (usually $5.5$) have $T_l$'s too
 long to be safely determined from unrenormalized integrations of at
 least 200,000 $P_J$ and hence can be regarded as regular.
 \label{FigG}}
 \end{figure}
 
 \begin{figure}
 \caption{
 A sample of regular and chaotic orbits at the 3:2 mean-motion
 resonance. In ({\it a}) Jupiter and Saturn move in a planar 
 approximation to their present orbits, but in ({\it b}) their eccentricities
 only have been increased by a factor of two so that their average values become
 $0.088$ and $0.094$. These higher $e$'s (though only Jupiter's is important)
 have the effect of broadening and deepening the 2/1 secondary resonance,
 as is shown in ({\it c}), but the enhancement is far less than at the 2:1
 mean-motion resonance.
 \label{FigH}}
 \end{figure}
 
 \begin{figure}
 \caption{
 Dynamical lifetime throughout the outer solar system.  At each
 semimajor axis bin, six test particles were started at different
 initial longitudes.  The solid curve marks the trace of the minimum
 time survived as a function of semimajor axis.  The points mark the
 survival times of the other particles, indicating the spread in
 dynamical lifetime. For reference,
 the semimajor axes of Jupiter, Saturn, Uranus, and Neptune are 5.2,
 9.5, 19.2, and 30.1~AU, respectively.
 \label{Figtp}}
 \end{figure}
 
 \begin{figure}
 \caption{
 The vertical axis shows the double logarithm of the final phase space
 separation of initial nearby test particles.  This is proportional to
 the estimated Lyapunov time.  The conversion to Lyapunov time is
 marked inside the left vertical axis.  Figure from Mikkola \& Innanen (1995).
 \label{mi95}}
 \end{figure}
 
 \begin{figure}
 \caption{
 Similar to figure~9, this shows the survival time versus semimajor
 axis in the inner planet region.  Here the vertical axis is linear
 rather than logarithmic.  The positions of the terrestrial planets
 are marked for reference. Figure from Evans \& Tabachnik (1999).
 \label{et99}}
 \end{figure}
 
 \begin{figure}
 \caption{
 The dynamical lifetime in the Kuiper belt as a function of semimajor
 axis and eccentricity, from Duncan et al (1995).  The lifetime is
 color-coded.  Long-lived regions can be seen at the locations of mean
 motion resonances with Neptune.
 \label{FigDLB}}
 \end{figure}
 
 \begin{figure}
 \caption{ The Lyapunov time $T_l$ of the system consisting of the four
 giant outer planets Jupiter, Saturn, Uranus, and Neptune. The
 semimajor axis of Uranus is varied around the best estimate of
 $a_U=19.2189$, keeping all other elements fixed. One can
 see the 2:1 resonance with Neptune from $19$ to $19.1$, the 7:1
 resonance with Jupiter at $19.18$, and numerous three-body resonances
 at $19.22$, $19.26$, $19.3$, and $19.34$.
 }
 \label{Fig_TLa}
 \end{figure}
 
 \begin{figure}
 \caption{ 
 An enlarged version of figure~\ref{Fig_TLa} around $19.22$ showing
 individual mean-motion resonances. 
 }
 \label{Fig_TLb}
 \end{figure}
 
 \begin{figure}
 \caption{ The phase space distance $ln[d(t)]$ between two slightly ($1mm$)
 displaced copies of the four giant planets. Two integrations are
 shown, one with $a_U=19.23$ and one with $a=19.26$. Both appear to
 follow power laws for the first 200~Myr, but one eventually shows
 rapid separation and hence is chaotic. The other appears to be regular.
 }
 \label{Fig_long}
 \end{figure}
 
 \begin{figure}
 \caption{ The resonant argument
 $3\lambda_J-5\lambda_S-7\lambda_U+3g_5t+6g_6t$. The libration period
 is about 20~Myr. One can see transitions from libration to rotation
 and back, eventually followed by a long period of rotation.
 }
 \label{Fig_jsu}
 \end{figure}

\begin{figure}
\caption{The ejection time, $T_c$, versus the Lyapunov time, $T_l$, in units of
the period of the test particle. About 90\% of the points fall within
a factor of 10 of the relation, $T_c \sim T_l^{1.75}$.}
 \label{Fig_tl_te}
\end {figure}

 
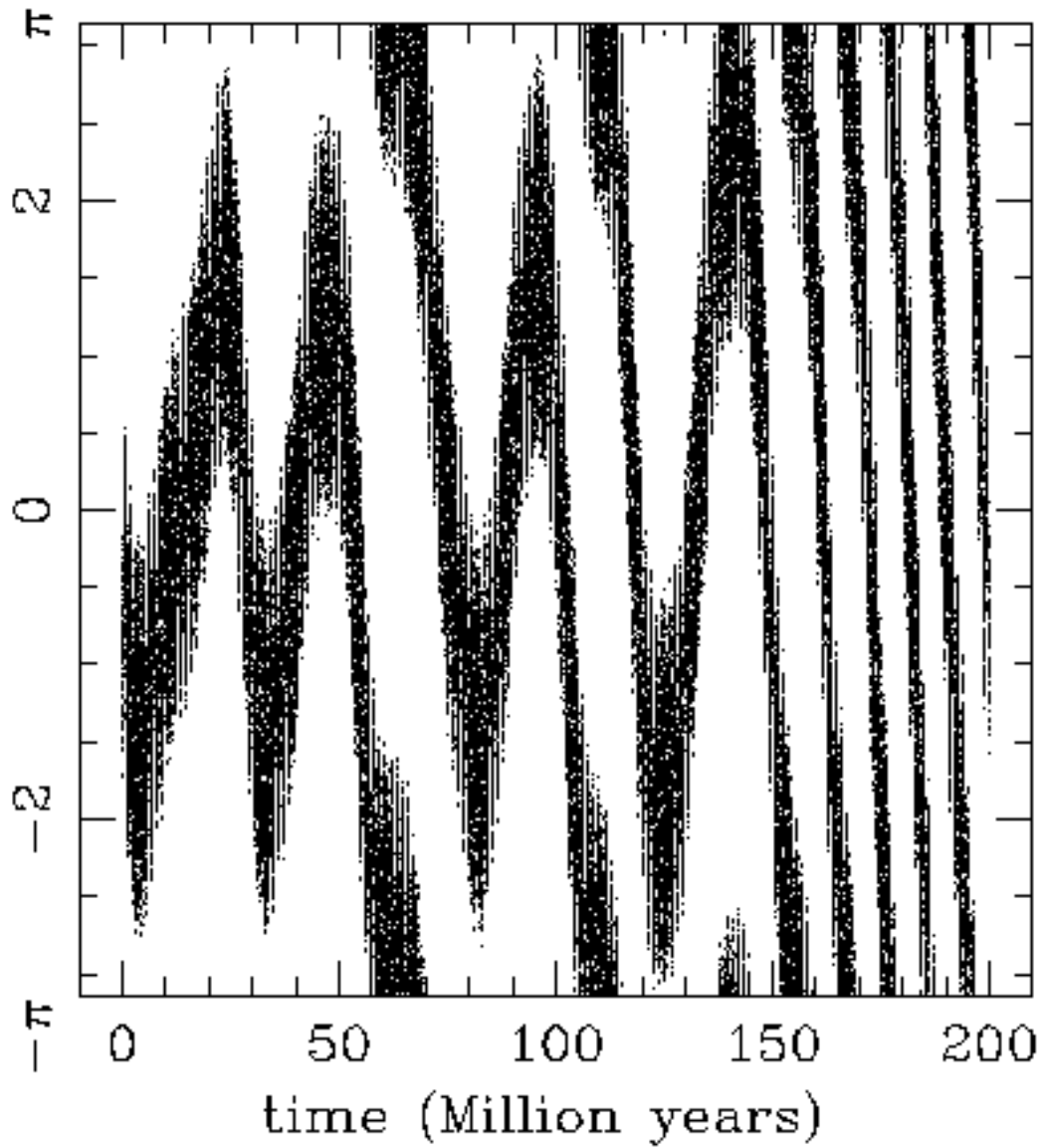
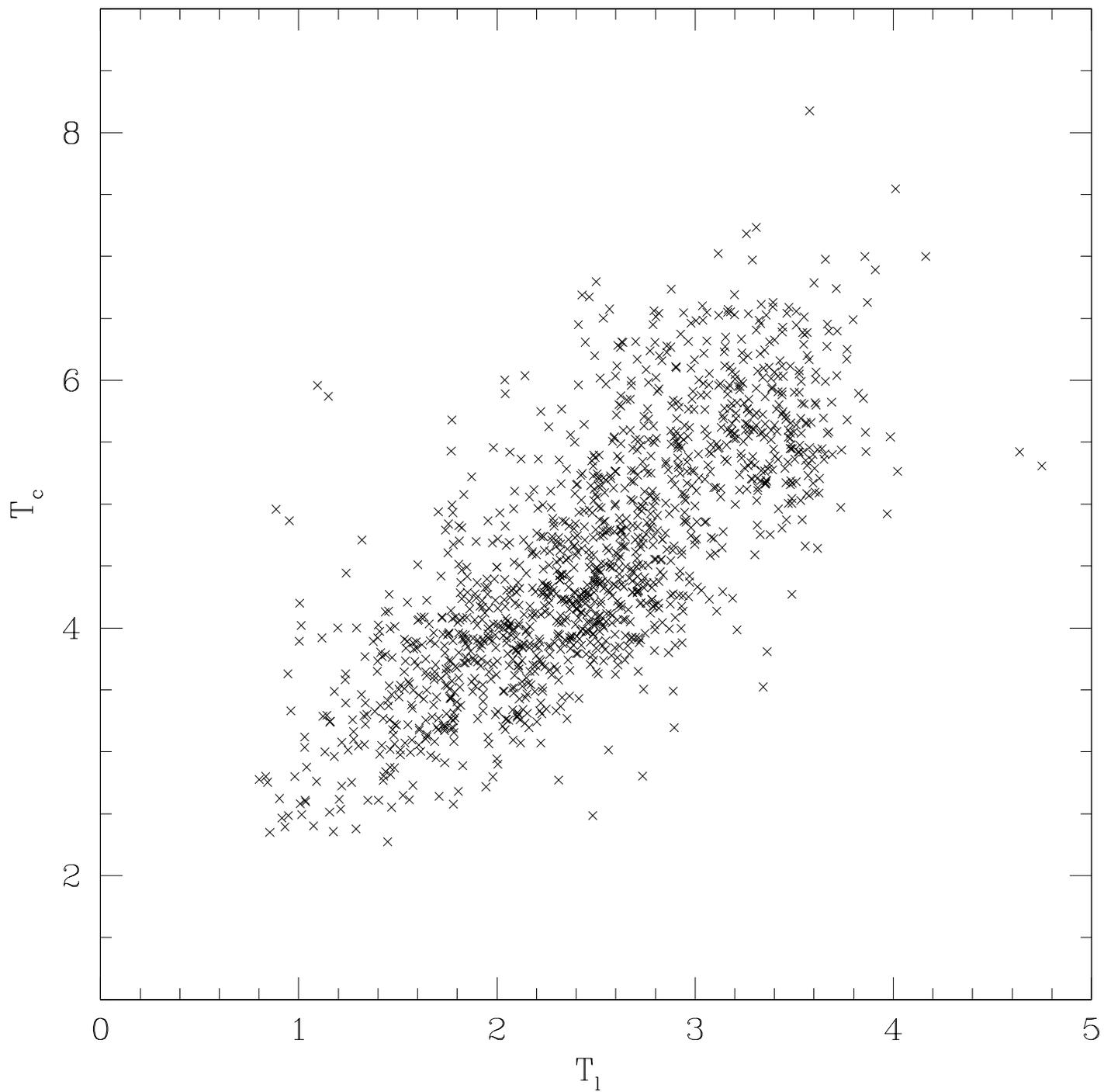
\end{document}